\documentclass{article}
\usepackage{amsfonts,amssymb, amsmath}
\usepackage[english]{babel}

\def\nn{\nonumber }
\def\bq{ \begin{equation}}
\def\eq{ \end{equation}}
\def\ben{ \begin{eqnarray}}
\def\en{ \end{eqnarray}}

\newtheorem{prop}{Proposition}
\begin{document}


\title{Second order Killing tensors related to symmetric spaces }
\author{E.O.Porubov, A.V.Tsiganov\\
\it\small St.Petersburg State University, St.Petersburg, Russia\\
\it\small e--mail: evg.porub@gmail.com, andrey.tsiganov@gmail.com}
\date{}
\maketitle

\begin{abstract}
We discuss the pairs of quadratic integrals of motion belonging to the $n$-dimensional space of independent integrals of motion in involution, that provide integrability of the corresponding Hamiltonian equations of motion by quadratures. In contrast to the Eisenhart theory, additional integrals of motion are polynomials of the fourth, sixth and other orders in momenta. The main focus is on the second-order Killing tensors corresponding to quadratic integrals of motion and relating to the special combinations of rotations and translations in Euclidean space.

 \end{abstract}


\section{Introduction}

\setcounter{equation}{0}
Let $A$ and $B$ be non-degenerate symmetric second-order tensor fields on Euclidean space $\mathbb R^n$. If the Schouten bracket between them is zero
\[[\![A,B]\!]=0\]
and the eigenvalue problem
\bq\label{eigen}
(A-\lambda B)\psi=0
\eq
has $n$ simple real eigenvalues and normal eigenvectors, then $A$ and $B$ generate a $n$-dimensional linear space of second-order tensor fields,
all in involution and with common eigenvectors. It allows us to calculate $n$ independent functions on the cotangent bundle $T^*\mathbb R^n$
\[
T_1=\sum_{ij} A^{ij}p_ip_j\,,\quad T_2=\sum_{ij} B^{ij}p_ip_j\,,\quad T_3=\sum_{ij} K_3^{ij}p_ip_j,\quad\ldots,\quad T_n=\sum_{ij} K_n^{ij}p_ip_j
\]
in involution
\[
\{T_i,T_j\}=0\,,\]
with respect to canonical Poisson brackets
\[\{q_i,q_j\}=0\,,\qquad \{p_i,p_j\}=0\,,\qquad \{q_i,p_j\}=\delta_{ij}\,,\quad i,j=1,\ldots,n\,.
\]
By adding suitable potentials
\[
H_1=T_1+V_1(q_1,\ldots,q_n)\,,\quad H_2=T_2+V_2(q_1,\ldots,q_n),\quad\ldots,\quad H_n=T_n+V_n(q_1,\ldots,q_n)
\]
we obtain the $n$-dimensional space of first integrals in involution \cite{eis34}, see also \cite{ben16,ts05,smir05,kal80}.

Thus, second-order tensors $A$ and $B$ define the completely integrable system, if they satisfy a set of conditions in $\mathbb R^n$ which can be verified without an explicit calculation of all the integrals of motion.

In \cite{ts15,ts15a,ts22a,ts22b} we found a few pairs of the second-order tensors $A$ and $B$ which also defines integrable and superintegrable systems on $T^*\mathbb R^n$, but the corresponding eigenvalue problem (\ref{eigen}) has no simple eigenvalues and normal eigenvectors. Our main aim is to construct enough examples to find the new criterion that two tensors $A$ and $B$ in $\mathbb R^n$ define a completely integrable system on $T^*\mathbb R^n$. It is natural to say that these tensors $A$ and $B$ are completely integrable.

In this note, we consider tensors $A$ and $B$ related to the special linear combinations of rotation and translations in Euclidean space $\mathbb R^n$. They are associated with Newton's equations of motion
\bq\label{eqm}
 \ddot{q}^\alpha=\sum_{\beta,\gamma,\delta} \mathcal R^\alpha_{\beta,\gamma,-\delta}q^\beta q^\gamma q_\delta-\omega_\alpha q_\alpha,\qquad \alpha,\beta,\gamma,\delta=1,\ldots, N
 \eq
and Hamiltonian
\bq\label{ham-gen}
H=\frac12\sum_{\alpha}\mathrm g^{\alpha,-\alpha}p_\alpha^2-\frac14\sum_{\alpha,\beta,\gamma,\delta} \mathcal R_{-\alpha,\beta,\gamma,-\delta}q^\alpha q^\beta q^\gamma q^\delta
+\frac12\sum_\alpha \omega_\alpha \left(q^\alpha\right)^2
\eq
 which were studied in \cite{f83, f86, r86}. Here $\mathrm g^{\alpha,-\alpha}$ and $R^\alpha_{\beta,\gamma,-\delta}$ are metric and curvature tensors (constant tensors), which correspond to classes A.III, BD.I, C.I and D.III of symmetric spaces in the Cartan classification.

 In this case, $A=\mathrm g$ is metric in Euclidean space and $B=K$ is a Killing tensor, which satisfies the Killing equation
\bq\label{kill-eq}
\nabla_iK^{jk} + \nabla_jK^{ki} + \nabla_kK^{ij} = 0,
\eq
where $\nabla$ is the Levi-Civita connection of $\mathrm g$.

 Relation of these Hamiltonian systems with the generalised multicomponent NLS hierarchy gives the Lax matrix
 \bq\label{lax-gen}
 L(\lambda)=\lambda^2 \mathcal A+\lambda\sum_{\alpha} q^\alpha\bigl(e_\alpha-e_{-\alpha}\bigr)-\frac{1}{a}\sum_{\alpha} \mathrm g^{\alpha,-\alpha}p_\alpha\bigl(e_\alpha+e_{-\alpha}\bigr)
 +\frac{1}{a}\sum_{\alpha,\beta}q_\alpha q_\beta[e_\alpha,e_{-\beta}]+\Lambda
 \eq
 and integrals of motion in the involution. For instance, Hamiltonian (\ref{ham-gen}) is equal to
 \[
H=\left.\frac{1}{4}\mbox{tr}\,L^2(\lambda)\right|_{\lambda=0}\,, 
 \]
because metric and curvature tensor are defined  using Weil generators
\[
\mathrm g^{\alpha,\beta}=\langle e_\alpha,e_\beta\rangle\,,
\quad\mbox{and}\quad\mathcal R_{\alpha,\beta, \gamma, \delta}=\langle [ e_{\alpha}, e_{\beta}], [ e_{\gamma}, e_{\delta}]\rangle \]
and Killing form $\langle X,Y\rangle=b\,\mbox{tr}(X\cdot Y$, $b\in \mathbb R$.

This Hamiltonian  $H$ (\ref{ham-gen}) is in involution with the polynomial of fourth order in momenta
\bq\label{G4}
G=\left.\mbox{tr}\,\right|_{\lambda=0}L^4(\lambda)=-\frac14\sum_{\alpha,\beta,\gamma,\delta} \mathcal R_{-\alpha,\beta,\gamma,-\delta}p^\alpha p^\beta p^\gamma p^\delta+\sum_{\alpha,\beta} S^{\alpha,\beta}
(q)p_\alpha p_\beta+W(q)\,,
\eq
whose leading part is defined by the curvature tensor $\mathcal R$.

Matrix  $\Lambda$  in (ref{lax-gen}) depends on parameters $\omega_i$ and describes a shift of the orbit obtained in the framework of the  Adler-Kostant-Symes theorem \cite{r86}. All these results were reproduced in the various textbooks \cite{pe90,rs94, rs03, tf94}, where readers can find all the necessary definitions and details, see also definitions of the Cartan involution and Weil generators in \cite{ger21}.

When all $\omega_\alpha=0$ and $\Lambda=0$, Hamiltonian $H$ (\ref{ham-gen}) commutes with a family of the noncommutative linear integrals of motion associated with the various combinations of rotations. In this case,  spectral invariants of the Lax matrix allow us to find a family of commuting integrals of motion, the numbers of which do not permit us to talk about integrability by the Liouville theorem in the general case.

If $\omega_\alpha\neq 0$ spectral invariants of the the Lax matrix $L(\lambda)$ (\ref{lax-gen}) generate a necessary number of the integrals of motion, which are polynomials in momenta of order two, four, six, eight, etc. The leading parts of these polynomials
,\[
H_i^{(2\ell)}=\sum_{jk\ldots m}^{2\ell} K_i^{jk\ldots m} p_jp_k\cdots p_m+\cdots\,,\qquad \ell=1,2,\ldots,
\]
 define the Killing tensors of valency $2\ell$ in Euclidean space $\mathbb R^n$
 \[
 [\![g,K_i]\!]=0\,,
 \]
 where $[\![.,.]\!]$ is a Schouten bracket. Below we will restrict ourselves to the study of second-order Killing tensors in a few low-dimensional Euclidean spaces.

\subsection{Killing tensors of valency two}
In Euclidean space, the generic Killing tensor of valency two is given by
 \bq \label{kill2-gen}
 K=\sum_{i,j} a_{ij} X_i\circ X_j+\sum_{i,j,k} b_{ijk} X_i\circ X_{j,k}+\sum_{i,j,k,m} c_{ijkm} X_{i,j}\circ X_{k,m}\,,
 \eq
 where
 \[
 X_i=\partial _i \qquad X_{i,j}=q_iX_j-q_jX_i\,,\qquad \partial_k=\frac{\partial}{\partial q_k}
 \]
is a basis of translations and rotations, $a_{ij}, b_{ijk}$ and $c_{ijm}$ are parameters and $\circ$ denotes symmetric product. To describe the rotation basis, we will use, as usual, the angular momentum tensor $J$ with components
\[
J_{ik}=q_ip_k-q_kp_i\,.
\]

 Dimension of vector space of the Killing tensors of valency $m$ in $n$-dimensional Euclidean space is given by the Delong-Takeuchi-Thompson formula
\[
d=\frac{1}{n}\binom{n+m}{m+1}\binom{n+m-1}{m}=\frac{1}{n}\binom{n+2}{3}\binom{n+1}{2}=\frac{n(n + 2)(n + 1)^2}{12}\,,
\]
where we put $m=2$ calculating the number of parameters $a_{ij}, b_{ijk}$ and $c_{ijkm}$.

Thus, we can find all the Killing tensors of valency two related to Hamiltonian $H=T+V$ (\ref{ham-gen}) solving the equation
\bq\label{dkdv}
d\left(KdV\right)=0\,,
\eq
which means that 1-form $KdV$ is an exact. Here $V$ is a function on $\mathbb R^n$, canonically lifted to $T^*\mathbb R^n$ and $KdV$ denotes
the 1-form image of $dV$ by $K$, interpreted as a linear endomorphism over 1-forms, whose components are $g_{\alpha,\beta} K^{\beta,\gamma} \partial_\gamma V$.

Substituting $K$ (\ref{kill2-gen}) and potential
\[
V=\frac14\sum_{\alpha,\beta,\gamma,\delta} \mathcal R_{-\alpha,\beta,\gamma,-\delta}q^\alpha q^\beta q^\gamma q^\delta
-\frac12\sum_\alpha \omega_\alpha \left(q^\alpha\right)^2
\]
into (\ref{dkdv}) we obtain a linear system of equations for coefficients $a_{ij}$, $b_{ijk}$ and $c_{ijkm}$ which can be solved on a computer. Then we can study the properties of the obtained solutions. Construction of the polynomials of higher order in momenta commuting with the Hamiltonian $H$ is an open question in the frameworks of Euclidean geometry.

According to \cite{ben16,eis34,smir05} Sylvester's criterion can be used to verify that $K$ has real and simple eigenvalues with respect to metric $\mathrm g$. The vanishing of the Haantjes torsion of $K$ allows us to prove that eigenvectors are normal, see historical remarks in \cite{smir05}.

The Haantjes tensor (torsion)
\[
 H_K(u,v)=K^2N_K(u,v)+N_K(Ku,Kv)-K\left( N_K(Ku,v)+N_K(u,Kv) \right)\,,
\]
is defined by using the Nijenhuis tensor (torsion)
\[
N_K(u,v)= K^2[u,v]+[Ku,Kv]-K\left([Ku,v]+[u,Kv]\right)\,,
\]
where $u,v$ are arbitrary vector fields and $ [. ,. ]$ denotes the commutator of two vector fields \cite{smir05}.

Below we prove that associated with Hamiltonian $H$ (\ref{ham-gen}) Killing tensors of valency two have non-zero Haantjes torsion. Nevertheless,
using Lax matrix $L(\lambda)$ (\ref{lax-gen}) we can get a set of completely integrable Killing tensors related to the special values of parameters $a_{ij}, b_{ijk}$ and $c_{ijkm}$ in (\ref{kill2-gen}). It could be useful to construct similar Killing tensors on non-Euclidean space \cite{eis35,ts22a,ts22b,pen70}.

\section{Symmetric spaces of A.III type}
\setcounter{equation}{0}
Let us consider equations of motion (\ref{eqm}) in Euclidean space $ \mathbb R^{mn}$ and Hamiltonian $H$ (\ref{ham-gen})
associated with the Riemannian pair
\[SU(m+n)/S\bigl(U(m)\times U(n)\bigr)\,,\qquad m\leq n\,,\qquad n+m\geq 4\,.\]
The typical representation of $su(m+n)$ is a set of $(m+n)\times (m+n)$ matrices with an obvious block-matrix
 structure associated with the following Cartan decomposition
 \[\mathfrak{g} \equiv \mathfrak k\oplus \mathfrak p,\qquad \mathfrak k=s(u(m) \oplus u(n))\,.\]
Here $\mathfrak k$ consists of block-diagonal matrices, while the linear space $\mathfrak p$ is spanned by block-off-diagonal matrices:
\[
\mathfrak k \simeq \left(\begin{array}{cc} u(m) & 0\\ 0 & u(n)   \end{array}\right), \qquad \mathfrak p \simeq \left(\begin{array}{cc} 0  &  P+\mathrm iQ \\ P^T-\mathrm i Q^T & 0   \end{array}\right).
\]
The corresponding Cartan element $A$ in (\ref{lax-gen}) acts as $-I$ on $so(m)$ and as $I$ on $so(n)$.

Following to \cite{f83,f86} we choose a representation in which Lax matrix (\ref{lax-gen}) reads as
 \bq\label{lax-a}
L(\lambda)=
\left(
 \begin{array}{cc}
 -2\lambda^2 I_m +QQ^T+a&0 \\ \\
 0 & 2\lambda^2 I_n-Q^TQ+b\\
 \end{array}
 \right)+\left(
 \begin{array}{cc}
 0 & P-2\mathrm{i}\lambda Q \\ \\
 P^T+2\mathrm{i}\lambda Q^T & 0\\
 \end{array}
 \right)\,,
 \eq
 where $I_m$ and $I_n$ are the $m\times m$ and $n\times n$ unit matrices,
 $a$ and $b$ are diagonal matrices depending on $m$ real numbers $a_k$ and $n$ real numbers parameters $b_i$
 \[
a=\mbox{diag}_m(a_1,\ldots,a_m)\,,\qquad b=\mbox{diag}_n(b_1,\ldots,b_n)\,,\qquad a_i,b_i\in \mathbb R\,,
\]
and $T$ means matrix transposition.

Matrices $P$ and $Q$ are $m\times n$ matrices depending linearly on $p$ and $q$ with entries
\[P_{ij}=p_{(i-1)n+j}\,,\qquad\mbox{and}\qquad Q_{ij}=q_{(i-1)n+j}\,,\qquad i=1,\ldots,m,\quad j=1,\ldots,n\,,\]
see examples below.

The Hamiltonian (\ref{ham-gen}) is given by
\begin{align}\label{ham-gena}
H=&\left.\frac14 \mbox{tr\,} L^2\,\right|_{\lambda=0}-\frac14\sum_{j=1}^ m a_j^2-\frac{1}{4}\sum_{i=1}^n b_i^2= \frac12\sum_{i=1}^n p_i^2+\frac12\sum_{j=0}^{m-1}\left(\sum_{i=1}^n q_{jn+i}^2\right)^2\\
\nn\\
+&\sum_{k,j=0; k>j}^{m-1}\left(\sum_{i=1}^n q_{jn+i}q_{kn+i}\right)^2
+\frac12\sum_{j=0}^{m-1}a_{j+1}\left(\sum_{i=1}^n
q_{jn+i}^2\right)
-\frac12\sum_{i=1}^n b_i\left(
\sum_{j=0}^{m-1}q_{jn+i}^2\right)\,.
\nn
\end{align}
When $a_i\neq 0$ and $b_i\neq 0$, there are two basic sets of integrals of motion obtained
from the characteristic polynomial of the Lax matrix
\[
\tau(z,\lambda)=\det\bigl(z\,I-L(\lambda)\bigr)\,,
\]
which are associated with $so(m)$ and $so(n)$, respectively. Because
\[
\{\tau(x,\lambda),\tau(y,\lambda)\}=0\,,
\]
all these integrals of motion are in the involution for each other.

\subsection{First set of the independent integrals of motion}
 The $m$ residues of the function
 \bq\label{delta1}
 \Delta_1(z,\lambda)=\frac{\tau(z,\lambda)}{\prod_{i=1}^m (z-a_i+2\lambda^2)}
 \eq
 at $z=a_i-2\lambda^2$ generate $mn$ independent integrals of motion $h_i^{(2\ell)}$
 \[
 \left.\mbox{Res}\,\Delta_1(z,\lambda)\right|_{z=a_i-2\lambda^2}=\sum_{k=0}^{n-1} \lambda^{2k} h^{\bigl( 2(n-k)\bigr)}_i\,,\qquad i=1,\ldots,m,
 \]
 which are polynomials of degree at most $2m$ since we take $m\leq n$. So, there are
 \begin{itemize}
 \item $m$ quadratic polynomials in momenta $h_1^{(2)},\ldots, h_m^{(2)}$;
 \item $m$ quartic polynomials in momenta $h_1^{(4)},\ldots, h_m^{(4)}$;
 \item $m$ sextic polynomials in momenta $h_1^{(6)},\ldots, h_m^{(6)}$;
 \item \begin{minipage}{8cm} \dotfill \end{minipage}
 \item $m$ polynomials of $2m$-order in momenta $h_1^{(2m)},\ldots, h_m^{(2m)}$
 \end{itemize}
and $m(n-m)$ remaining polynomials of $2m$-order in momenta.

Polynomials of the second order in momenta have the following form
\bq\label{fm-gen}
h_i^{(2)}=\sum_{k\neq i}^m \frac{M_{ik}^2}{a_i-a_k} + t_i(p)+v_i(q)\,,
\eq
where functions
\[M_{ik}=\sum^n J_{ j\ell}\,,\qquad J_{j\ell}=q_jp_\ell-q_\ell p_j\,,\]
constitute realization of Lie algebra $so^*(m)$ associated with compositions of $n$ simple rotations in $ \mathbb R^{mn}$.

Functions $t_i(p)$ correspond to compositions of the $n$ translations
\[
t_i(p)=\sum^n p_\ell^2\,,
\]
and $v_i(q)$ are polynomials of the fourth order in coordinates $q_i$.

\subsection{Second set of the independent integrals of motion}
The $n$ residues of the function
 \bq\label{delta2}
 \Delta_2(z,\lambda)=\frac{\tau(z,\lambda)}{\prod_{i=1}^n (z-b_i-2\lambda^2)}
 \eq
 at $z=b_i+2\lambda^2$ generate $mn$ independent integrals of motion $H_i^{(2\ell)}$
 \[
 \left.\mbox{Res}\,\Delta_2(z,\lambda)\right|_{z=b_i+2\lambda^2}=\sum_{k=0}^{m-1} \lambda^{2k} H^{\bigl( 2(m-k)\bigr)}_i\,,\qquad i=1,\ldots,n
 \]
which are polynomials of order $2\ell$ in momenta. So, there are
 \begin{itemize}
 \item $n$ quadratic polynomials in momenta $H_1^{(2)},\ldots, H_n^{(2)}$;
 \item $n$ quartic polynomials in momenta $H_1^{(4)},\ldots, H_n^{(4)}$;
 \item $n$ sextic polynomials in momenta $H_1^{(6)},\ldots, H_n^{(6)}$;
 \item \begin{minipage}{8cm} \dotfill \end{minipage}
 \item $n$ polynomials of $2m$-order in momenta $H_1^{(2m)},\ldots, H_n^{(2m)}$\,.
 \end{itemize}
In this case polynomials of second order in momenta have the following form
\bq\label{fn-gen}
H_i^{(2)}= \sum_{k\neq i}^n \frac{N_{ik}^2}{b_i-b_k} +T_i(p)+U_i(q)\,,
\eq
where functions
\[N_{ik}=\sum^m J_{ j\ell}\,,\qquad J_{j\ell}=q_jp_\ell-q_\ell p_j\,,\]
form realization of $so^*(n)$ via compositions of $m$ simple rotations in $ \mathbb R^{mn}$.

Functions $T_i(p)$ correspond to compositions of the $m$ translations
\[
T_i(p)=\sum^m_\ell p_\ell^2\,,
\]
and $U_i(q)$ are polynomials of the fourth order in coordinates.

Summing up, we have $n+m-1$ quadratic integrals of motion
\[
h_1^{(2)}+\cdots+h_m^{(2)}=2H=H_1^{(2)}+\cdots+H_n^{(2)}\,,
\]
associated with the linear combinations of rotations, which realise $so^*(m)$ and $so^*(n)$, and with the linear combinations of translations.
\begin{prop}
Equations of motion (\ref{eqm}) defined by $H$ (\ref{ham-gena}) have only $n+m-1$ independent quadratic integrals of motion in involution.
\end{prop}
We can directly prove this proposition for low-dimensional case $\mathbb R^4$ substituting generic solution (\ref{kill2-gen}) of the Killing equation (\ref{kill-eq})
into (\ref{dkdv}) and solving the resulting system of linear equations. In the generic case, we have to calculate a number of unknown coefficients by the Delong-Takeuchi-Thompson formula and compare this number with a rank of this system of linear equations.

\subsection{Euclidean space $ \mathbb R^n$, case $m=1$}

When $m=1$ we have the so-called Garnier system and all the second-order Killing tensors
\[
K_i=-\sum _{k\neq i}\frac{ X_{ik}\cdot X_{ik}}{b_i-b_k}-X_i\cdot X_i
\]
consist of single rotation and single translation
\[X_{i,k}=q_i\partial_k-q_k\partial_i\,,\qquad X_i=\partial_i\,, \]
and their Haanties torsion is equal to zero. Thus, the Hamilton-Jacoby equation $H=E_i$ admits additive separation of variables. Separated variables are the standard elliptic coordinates in $ \mathbb R^n$ and, therefore, this integrable system constrained to an ellipsoid remains integrable \cite{st85}.

When $m>1$ the corresponding Killing tensors of valency two have nontrivial Haantjes torsion. It means that the Hamilton-Jacobi equation $H=E$ does not admit the separation of variables in the curvilinear orthogonal coordinates. Below we present a few examples of the corresponding quadratic integrals of motion.

\subsection{Euclidean space $ \mathbb R^4$, case $m=n=2$}
Because $so(4)\simeq so(2)\times so(2)$ there appear double rotations or Clifford displacements in $\mathbb R^4$, which can be associated with the left- and right-multiplication by a unit quaternion. It is a classical object in the geometry of the fourth-dimensional Euclidean space \cite{con03,pert03,man14}.

The $4\times 4$ Lax matrix (\ref{lax-a}) is equal to
 \bq\label{lax-so4}
 L(\lambda)=
\left(
 \begin{smallmatrix}
 q_1^2 + q_2^2 + a_1-2\lambda^2 & q_1q_3 + q_2q_4 & p_1 - 2\mathrm{i}\lambda q_1 & p_2 - 2\mathrm{i}\lambda q_2 \\ \\
 q_1q_3 + q_2q_4 & q_3^2 + q_4^2 + a_2-2\lambda^2 & p_3 - 2\mathrm{i}\lambda q_3 & p_4 - 2\mathrm{i}\lambda q_4 \\ \\
 p_1 - 2\mathrm{i}\lambda q_1& p_3 - 2\mathrm{i}\lambda q_3 & b_1-q_1^2 - q_3^2+2\lambda^2 & -q_1q_2 - q_3q_4 \\ \\
 p_2 - 2\mathrm{i}\lambda q_2& p_4 - 2\mathrm{i}\lambda q_4 & -q_1q_2 - q_3q_4 & b_2- q_2^2 - q_4^2 +2\lambda^2 \\
 \end{smallmatrix}
\right)\,,
\eq
so Hamiltonian $H$ (\ref{ham-gena}) has the form
\begin{align}\label{ham-13b}
H=&\frac{p_1^2}{2} +\frac{p_2^2}{2} +\frac{p_3^2}{2}+\frac{p_4^2}{2}
+\frac12(q_1^2 + q_2^2)^2 + \frac12(q_3^2 + q_4^2)^2 + (q_1q_3 + q_2q_4)^2
\\
\nn\\
+&\frac{a_1-b_1}{2}\,q_1^2 +\frac{a_1-b_2}{2}\,q_2^2 + \frac{a_2-b_1}2\,q_3^2 +\frac{a_2-b_2}{2}\, q_4^2 \,.
\nn
\end{align}
Because
\[
\frac12(q_1^2 + q_2^2)^2 + \frac12(q_3^2 + q_4^2)^2 + (q_1q_3 + q_2q_4)^2
 =\frac{(q_1^2 + q_2^2 + q_3^2+q_4^2)^2}{2}
-(q_1q_4 - q_2q_3)^2
\]
this Hamiltonian coincides with (13b) case from the paper \cite{f86} after permutation of indexes.

Spectral curve of the Lax matrix $L(\lambda)$ (\ref{lax-so4}) is a non-hyperelliptic curve defined by characteristic equation
\[\mathcal C:\qquad\det\Bigl( zI-L(\lambda)\Bigr)=0.\] In generic case its
 genus is equal to five $g=5$, when $a_1=a_2$ or $b_1=b_2$ genus of this non-hyperelliptic curve $\mathcal C$ is equal to four $g=4$.

\vskip0.2truecm
\par\noindent\textbf{First set of integrals of motion}
\par\noindent
Two residues of the function $\Delta(z,\lambda)$ (\ref{delta1})
\[
\Delta(z,\lambda)=\frac{\det\Bigl(zI-L(\lambda)\Bigr)}{(z-a_1+2\lambda^2)(z-a_2+2\lambda^2)}
\]
are equal to
\[
\left.\mbox{Res}\right|_{z=a_i-2\lambda^2}\, \Delta(z,\lambda)=4\lambda^2 f_i+g_i\,,\qquad i=1,2.
\]
where $f_{1,2}$ and $g_{1,2}$ are the second and fourth-order polynomials in momenta, respectively.

Third residue in infinity
\[\left.\mbox{Res}\right|_{z=\infty}\, \Delta(z,\lambda)=-4\lambda^2 (f_1+f_2) -(g_1+g_2)\,,\]
give rise to integrals of motions
$ f_1 + f_2 =2H$ and $g_1+g_2=f_3$
which are polynomials of the second order in momenta.

Quadratic integrals of motion $f_{1,2}$ have the following form
\bq\label{f22}
f_1=-\frac{M_{12}^2}{a_1-a_2}+p_1^2+p_2^2+ v_1\qquad\mbox{and}\qquad f_2= \frac{M_{12}^2}{a_1-a_2}+p_3^2+p_4^2+v_2,
\eq
where
\begin{align*}
v_1=&(q_1^2 + q_2^2 + q_3^2 + a_1 - b_1)q_1^2 + (q_1^2+q_2^2 + q_4^2 + a_1 - b_2)q_2^2 + 2q_1q_2q_3q_4\,,\\
\\
v_2=&(q_1^2 + q_3^2 + q_4^2 + a_2 - b_1) q_3^2 + (q_2^2+q_3^2 + q_4^2 + a_2 - b_2) q_4^2 + 2q_1q_2q_3q_4\,.
\end{align*}
Here $M_{12}$ is a function associated a double rotation in $\mathbb R^4$
\[
M_{12}=J_{1,3}+J_{2,4}=(q_1p_3-q_3p_1)+(q_2p_4-q_4p_2)\,.
\]
It commutes with terms in $f_{1,2}$ associated with translations
\[
\{M_{12},p_1^2+p_2^2\}=\{M_{12},p_3^2+p_4^2\}=0\,,
\]
and with function associated with the second independent double rotation in $\mathbb R^4$
\[
N_{12}=J_{1,2}+J_{3,4}=( q_1p_2 -q_2p_1) + (q_3p_4 - p_3q_4)\,,
\]
so that
\[\{M_{12},N_{12}\}=0\,.
\]
The second independent rotation appears in the following linear combination of the integrals of motion
\begin{align*}
f_3=&(b_1 + b_2)H - g_1 - g_2 - a_1f_1 - a_2f_2\\
\\
=&N_{12}^2 -\scriptstyle \frac{(b_1-b_2)\left((q_1^2 + q_2^2 + q_3^2 + q_4^2)(q_1^2 - q_2^2 + q_3^2 - q_4^2)+(q_1^2-q_2^2)a_1 + (q_3^2-q_4^2)a_2- (q_1^2+q_3^2)b_1 + (q_2^2+q_4^2)b_2\right)}2.
\end{align*}
 When $b_1=b_2$ linear integral of motion $N_{12}$ is a function on $f_{1,2}$ and $g_{1,2}$.

The leading term in quartic polynomials in momenta is the perfect square
\[
(a_1-a_2)g_1=(p_1p_4 - p_2p_3)^2+\cdots\qquad\mbox{and}\qquad (a_2-a_1)g_1=(p_1p_4 - p_2p_3)^2+\cdots\,.
\]
There is also a combination of the integrals of motion
\[
g_3=2H^2 - a_1g_1 - a_2g_2
\]
with the leading part defined by the curvature tensor $\mathcal R$ (\ref{ham-gen}, \ref{ham-13b})
\[
g_3=-\frac14\sum_{\alpha,\beta,\gamma,\delta} \mathcal R_{-\alpha,\beta,\gamma,-\delta}p^\alpha p^\beta p^\gamma p^\delta+\cdots
=-\frac12 (p_1^2 + p_2^2 + p_3^2 + p_4^2)^2 - (p_1p_4 - p_2p_3)^2+\cdots.
\]

\vskip0.2truecm
\par\noindent\textbf{Second set of integrals of motion}
\par\noindent
Residues of the function $\Delta(z,\lambda)$ (\ref{delta2})
\[
\Delta(z,\lambda)=\frac{\det\Bigl(zI-L(\lambda)\Bigr)}{(z-b_1-2\lambda^2)(z-b_2-2\lambda^2)}
\]
are equal to
\begin{align*}
\left.\mbox{Res}\right|_{z=b_i+2\lambda^2}\, \Delta(z,\lambda)=-4\lambda^2 F_i+G_i\,,\qquad i=1,2.
\\
\\
\left.\mbox{Res}\right|_{z=\infty}\, \Delta(z,\lambda)=8\lambda^2 H -(G_1+G_2)\,,\qquad F_1 + F_2 - 2H=0\,,
\end{align*}
where polynomials of second order in momenta $F_{1,2}$ and $G_1+G_2$ are independent for each other.

Quadratic integrals of motion $F_{1,2}$ are equal to
\[
F_1=\phantom{-}\frac{N_{12}^2}{b_1-b_2} + p_1^2 + p_3^2+V_1\qquad\mbox{and}\qquad
F_2=-\frac{N_{12}^2}{b_1-b_2} + p_2^2 + p_4^2+V_2\,,\
\]
where
\[\begin{array}{rcl}
V_1&=&(q_1^2+q_2^2 + q_3^2 +a_1- b_1)q_1^2 + (q_1^2+q_3^2 + q_4^2+a_2- b_1)q_3^2 + 2q_1q_2q_3q_4\,,\\
\\
V_2&=&(q_1^2+q_2^2 + q_4^2 +a_1- b_2)q_2^2 + (q_2^2+q_3^2 + q_4^2 +a_2- b_2)q_4^2+2q_1q_2q_3q_4\,.
\end{array}
\]
Here $N_{12}$ is a function associated with the double rotation in $\mathbb R^4$:
\[
N_{12}=J_{1,2}+J_{3,4}=( q_1p_2 -q_2p_1) + (q_3p_4 - p_3q_4)\,.
\]
This function commutes with terms in $F_{1,2}$ associated with translations
\[\{N_{12},p_1^2+p_3^2\}=\{N_{12},p_2^2+p_4^2\}=0\,,\]
and with function associated with the independent second double rotation
\[
M_{12}=J_{1,3}+J_{2,4}=( q_1p_3 -q_3p_1) + (q_2p_4 - p_2q_4)\,,
\]
which was included in the definition of $f_{1,2}$ (\ref{f22}) from the first set of integrals of motion.

This function also appears in the following linear combination of integral of motion from the second set of integrals of motion
\begin{align*}
F_3&=G_1+G_2-b_1F_1-b_2F_2-(a_1 + a_2)H\\
\\
&=M_{12}^2+\scriptstyle\frac{(a_1 - a_2)\bigl(
p_3^2 + p_4^2 - p_1^2 - p_2^2 + (q_1^2 - q_3^2)b_1 + (q_2^2 - q_4^2)b_2 - (q_1^2 + q_2^2)a_1 + (q_3^2 + q_4^2)a_2 - (q_1^2 + q_2^2)^2 + (q_3^2 + q_4^2)^2
\bigr)}{2}\,.
\end{align*}
When $a_1=a_2$ linear integral of motion $N_{13}$ is a function on $F_{1,2}$ and $G_{1,2}$.

The leading term in the quartic invariants is a perfect square
\[(b_1-b_2)G_{1}=(p_1p_4-p_2p_3)^2+\ldots\qquad\mbox{and}\qquad (b_2-b_1)G_{2}=(p_1p_4-p_2p_3)^2+\ldots,.\]
As above, there are quartic invariant \[G_3=2H^2-b_1G_1-b_2G_2\] with leading term defined by the curvature tensor $\mathcal R$ (\ref{ham-gen}, \ref{ham-13b})
\[
G_3=-\frac14\sum_{\alpha,\beta,\gamma,\delta}\mathcal R_{-\alpha,\beta,\gamma,-\delta}p^\alpha p^\beta p^\gamma p^\delta+\cdots
=\frac12 (p_1^2 + p_2^2 + p_3^2 + p_4^2)^2 - (p_1p_4 - p_2p_3)^2+\cdots.
\]

Summing up, there are only  $m+n-1=3$ independent quadratic integrals of motion among $f_1,f_2,f_3$ and $F_1,F_2,F_3$ in $\mathbb R^4$. The corresponding Killing tensors of valency two have non-zero Haantjes torsion.

\subsection{Euclidean space $ \mathbb R^6$, case $m=2$ and $n=3$}
The $5\times 5$ Lax matrix (\ref{lax-a}) reads as
 \[L(\lambda)=\left(
 \begin{smallmatrix}
 q_1^2 + q_2^2 +q_3^2+ a_1-2\lambda^2 & q_1q_4 + q_2q_5 + q_3q_6 & p_1 - 2\mathrm{i}\lambda q_1 & p_2 - 2\mathrm{i}\lambda q_2&p_3 - 2\mathrm{i}\lambda q_3 \\ \\
 q_1q_4 + q_2q_5 + q_3q_6 & q_4^2 + q_5^2 + q_6^2 + a_2-2\lambda^2 & p_4 - 2\mathrm{i}\lambda q_4 & p_5 - 2\mathrm{i}\lambda q_5&p_6 - 2\mathrm{i}\lambda q_6 \\ \\
 p_1 + 2\mathrm{i}\lambda q_1& p_4 + 2\mathrm{i}\lambda q_4 & b_1-q_1^2 - q_4^2+2\lambda^2 &-q_1q_2 - q_4q_5 &-q_1q_3 - q_4q_6\\ \\
 p_2 + 2\mathrm{i}\lambda q_2& p_5 + 2\mathrm{i}\lambda q_5 & -q_1q_2 - q_4q_5 & b_2- q_2^2 - q_5^2 +2\lambda^2&-q_2q_3 - q_5q_6 \\ \\
 p_3 + 2\mathrm{i}\lambda q_3& p_6 + 2\mathrm{i}\lambda q_6 &-q_1q_3 - q_4q_6 &-q_2q_3 - q_5q_6&b_3- q_3^2 - q_6^2 +2\lambda^2\\
 \end{smallmatrix}
 \right)\,,
\]
so Hamiltonian $H$ (\ref{ham-gen},\ref{ham-gena}) reads as
\begin{align}
H=&\frac12\,\displaystyle \sum_{i=1}^6 p_i^2+\frac{(q_1^2 + q_2^2 + q_3^2)^2}{2} +\frac{(q_4^2 + q_5^2 + q_6^2)^2}{2}+ (q_1q_4 + q_2q_5 + q_3q_6)^2\nn\\
\label{ham-13bb}\\
-&\dfrac{q_1^2+ q_4^2}{2}b_1 -\dfrac{q_2^2+ q_5^2}{2}b_2 -\dfrac{q_3^2+ q_6^2}{2}b_3
 +\dfrac{q_1^2 + q_2^2 + q_3^2}{2}a_1 +\dfrac{q_4^2 + q_5^2 + q_6^2}{2}a_2\,.\nn
\end{align}

When $a_i=0$ and $b_i=0$ this Hamiltonian commutes with the following linear integrals of motion
\begin{align*}
M_{12}=&(q_1p_4-p_4q_1) + (q_2p_5-p_2q_5) +(q_3p_6-p_3q_6)\,,\quad N_{12}= (q_1p_2-p_1q_2) + (q_4p_5-p_4q_5)\,,\nn\\
\nn\\
N_{13}=&(q_1p_3-p_1q_3)+(q_4p_6-p_4q_6)\,,\qquad N_{23}=(q_2p_3-p_2q_3)+(q_5p_6-p_5q_6)\,,
\end{align*}
associated with rotations in $\mathbb R^5$.

The equation for the spectral curve of the Lax matrix  contains only five commuting functions $H,F_1, F_2$ and $G_1, G_2$
\begin{align*}
 \tau(z,\lambda)=&z^5 - 2\lambda^2 z^4 - 2(4\lambda^4 + H)z^3 + (16\lambda^6 + 4H\lambda^2 + F_1)z^2\\ \\ +& (16\lambda^8 + 8H\lambda^4 - 4F_2^2\lambda^2 + G_1)z
  - 32\lambda^{10} - 16H\lambda^6 + (8F_2^2 - 4F_1)\lambda^4 - 2G_1\lambda^2 +G_2\,,
\end{align*}
where
\[
F_1=M_{12}^2-N_{12}^2-N_{13}^2-N_{23}^2\,,\qquad F_2=M_{12}^2\,.
\]
Thus, we must find the missing integral of motion using other properties of the Lax matrix
similar to the full Toda lattice \cite{deft86}.

In the generic case $a_i\neq 0$ and $b_i\neq0$ spectral curve of the Lax matrix $L(\lambda)$ is a genus six non-hyperelliptic curve, that allows us to get six independent integrals of motion in the involution.

\vskip0.2truecm
\par\noindent\textbf{First set of integrals of motion}
\par\noindent
Two residues of the function $\Delta(z,\lambda)$ (\ref{delta1})
\[
\Delta(z,\lambda)=\frac{\det\Bigl(zI-L(\lambda)\Bigr)}{(z-a_1+2\lambda^2)(z-a_2+2\lambda^2)}
\]
are equal to
\[
\left.\mbox{Res}\right|_{z=a_i-2\lambda^2}\, \Delta(z,\lambda)=-16\lambda^4 f_i+\lambda^2 g_i+w_i\,,\qquad i=1,2.
\]
where $f_{1,2}$ are polynomials of second order in momenta. Because $2m=4$ other integrals of motion $g_{1,2}$ and $w_{1,2}$ are polynomials of fourth order in momenta.

Residue at infinity is equal to
\[
\left.\mbox{Res}\right|_{z=\infty}\, \Delta(z,\lambda)=32\lambda^4 H -\lambda^2(g_1+g_2)-(w_1+w_2)\,,\qquad f_1 + f_2 - 2H=0\,,
\]
Integrals of motion $f_{1,2}$ are polynomials of second order in momenta
\bq\label{f23}
f_1=-\frac{M_{12}^2}{b_1-b_2}+p_1^2 + p_2^2 + p_3^2 +v_1\,,\quad\mbox{and}\quad
f_2=\frac{M_{12}^2}{b_1-b_2}+p_4^2 + p_5^2 + p_6^2 +v_2
\eq
where
\begin{align*}
v_1=&(q_1^2 + q_2^2 + q_3^2 + q_4^2 + a_1 - b_1 ) q_1^2 + (q_1^2 + q_2^2 + q_3^2 + q_5^2 + a_1 - b_2 ) q_2^2 \\
\\ +&(q_1^2 + q_2^2 + q_3^2 + q_6^2 + a_1 - b_3 ) q_3^2+ 2 q_1q_2 q_4 q_5 + 2q_1 q_3 q_4 q_6 + 2 q_2 q_3 q_5 q_6 \,,\\
\\
v_2=&(q_1^2 + q_4^2 + q_5^2 + q_6^2 + a_2 - b_1 ) q_4^2 + (q_2^2 + q_4^2 + q_5^2 + q_6^2 + a_2 - b_2 ) q_5^2\\ \\
+&(q_3^2 + q_4^2 + q_5^2 + q_6^2 + a_2 - b_3 ) q_6^2+ 2 q_1 q_2q_4 q_5 + 2 q_1 q_3q_4 q_6 + 2 q_2 q_3 q_5 q_6 \,.
\end{align*}
Here $M_{12}$ is a function associated with the triple rotation in $\mathbb R^6$:
\bq\label{3-rot}
M_{12}=J_{14}+J_{25}+J_{36}=(q_1p_4-p_4q_1) + (q_2p_5-p_2q_5) +(q_3p_6-p_3q_6)\,.
\eq
Linear combinations of other integrals of motion are associated with double rotations in $\mathbb R^6$. For instance, the polynomial of second order in momenta
\[
f_3=2(b_1 + b_2 + b_3)H + \frac{g_1 + g_2}{4} - 2a_1f_1 - 2a_2f_2\]
is equal to
\[
f_3=N_{12}^2 + N_{13}^2 + N_{23}^2+(p_1^2 + p_4^2)b_1 + (p_2^2 + p_5^2)b_2 + (p_3^2 + p_6^2)b_3+v_3\,,
\]
where
\begin{align}
N_{12}=& J_{12}+J_{45}=(q_1p_2-p_1q_2) + (q_4p_5-p_4q_5)\,,\nn\\
\nn\\
N_{13}=&J_{13}+J_{46}=(q_1p_3-p_1q_3)+(q_4p_6-p_4q_6)\,,\label{f-so3}\\
\nn\\
N_{23}=&J_{23} +J_{56}=(q_2p_3-p_2q_3)+(q_5p_6-p_5q_6)\,,\nn
\end{align}
and
\begin{align*}
v_3&= (q_1^4 + q_1^2 q_2^2 + q_1^2 q_3^2 + 2 q_1^2 q_4^2 + 2 q_1 q_2 q_4 q_5 + 2 q_1 q_3 q_4 q_6 + q_4^4 + q_4^2 q_5^2 + q_4^2 q_6^2 + a_1 q_1^2 + a_2 q_4^2) b_1\\
\\
 &+ (q_1^2 q_2^2 + 2 q_1 q_2 q_4 q_5 + q_2^4 + q_2^2 q_3^2 + 2 q_2^2 q_5^2 + 2 q_2 q_3 q_5 q_6 + q_4^2 q_5^2 + q_5^4 + q_5^2 q_6^2 + a_1 q_2^2 + a_2 q_5^2) b_2\\
 \\
&+ (q_1^2 q_3^2 + 2 q_1 q_3 q_4 q_6 + q_2^2 q_3^2 + 2 q_2 q_3 q_5 q_6 + q_3^4 + 2 q_3^2 q_6^2 + q_4^2 q_6^2 + q_5^2 q_6^2 + q_6^4 + a_1 q_3^2 + a_2 q_6^2) b_3\\
 \\
&-(q_1^2 + q_4^2) b_1^2- (q_2^2 + q_5^2) b_2^2-(q_3^2 + q_6^2) b_3^2\,.
\end{align*}
We will omit such explicit expressions for the bulky potentials below for brevity.
\vskip0.2truecm
\par\noindent\textbf{Second set of integrals of motion}
\par\noindent
Three residues of the function $\Delta(z,\lambda)$ (\ref{delta2})
\[
\Delta(z,\lambda)=\frac{\det\Bigl(zI-L(\lambda)\Bigr)}{(z-b_1-2\lambda^2)(z-b_2-2\lambda^2)(z-b_3-2\lambda^2)}
\]
 are equal to
\[
\left.\mbox{Res}\right|_{z=b_i+2\lambda^2}\, \Delta(z,\lambda)=4\lambda^2 F_i+G_i\,,\qquad i=1,2,3.
\]
where $F_i$ and $G_i$ are the second and fourth-order polynomials in momenta.

Residue in infinity reads as
\[
\left.\mbox{Res}\right|_{z=\infty}\, \Delta(z,\lambda)=8\lambda^2 H -(G_1+G_2+G_3)\,,\quad 2H+F_1+F_2+F_3=0\,.
\]
Polynomials of second order in momenta are defined by double rotations and double translations (\ref{fn-gen})
\begin{align*}
F_1&=-\dfrac{N_{12}^2}{b_1 - b_2} -\dfrac{ N_{13}^2}{b_1 - b_3}-p_1^2 - p_4^2
- (q_1^2+q_2^2 + q_3^2 + 2q_4^2 +a_1-b_1)q_1^2 \\
\\
&-( q_4^2 +q_5^2 + q_6^2+a_2- b_1)q_4^2-2 (q_2q_5 +q_3q_6)q_1q_4\,,
\\ \\
F_2&=-\dfrac{ N_{21}^2 }{b_2 - b_1} -\dfrac{N_{23}^2}{b_2 - b_3}-p_2^2-p_5^2
-(q_1^2+q_2^2 + q_3^2 + 2q_5^2 +a_1-b_2)q_2^2 \\
\\
&-( q_4^2+q_5^2+ q_6^2 + a_2 - b_2)q_5^2-2 (q_1q_4+q_3q_6)q_2q_5\,,
\\ \\
F_3&=-\dfrac{N_{31}^2}{b_3-b_1} -\dfrac{ N_{32} ^2}{b_3-b_2}-p_3^2-p_5^2-(q_1^2 + q_2^2+q_3^2 - 2q_6^2+a_1-b_3)q_3^2 \\
\\
&- (q_4^2 + q_5^2+q_6^2 + a_2 -b_3)q_6^2-2(q_1q_4 +q_2q_5)q_3q_6\,.
\end{align*}
Functions $N_{ij}=-N_{ji}$ (\ref{f-so3}) are associated with double rotations in $ \mathbb R^6$ and realisations of $so(3)$ algebra with the brackets
\[
\{N_{12}, N_{13}\} = N_{23}\,,\qquad \{N_{13}, N_{23}\}=N_{12}\,,\qquad
\{N_{23}, N_{12}\}=N_{13}\,.
\]
Leading term of the independent on $F_1,F_2$ and $F_3$ polynomial of second order in momenta
\begin{align*}
F_4&=G_1+G_2+G_3-b_1F_1-b_2F_2-b_3F_3-(a_1+a_2)H\\
\\
&=M_{12}^2-\frac{a_1-a_2}{2}\left(p_1^2 + p_2^2 + p_3^2-p_4^2 - p_5^2 - p_6^2+ V_4\right)
\end{align*}
includes function $M_{12}$ (\ref{3-rot}) associated with the triple rotation in $ \mathbb R^6$.
The corresponding potential $V_4$ is equal to
\begin{align*}
V_4&=(q_1^2 + q_2^2 + q_3^2 + q_4^2 + q_5^2 + q_6^2) (q_1^2 + q_2^2 + q_3^2 - q_4^2 - q_5^2 - q_6^2)\\
\\
&+
(q_1^2 + q_2^2 + q_3^2) a_1 - (q_4^2 + q_5^2 + q_6^2) a_2 - (q_1^2 - q_4^2) b_1 - (q_2^2 - q_5^2) b_2 - (q_3^2 - q_6^2) b_3\,.
\end{align*}
When $a_1=a_2$ linear polynomial $M_{12}$ commutes with all the integrals of motion $H$, $F_k$ and $G_k$.

The following combination of quartic integrals of motion
\[G_4=2H^2-b_1G_1-b_2G_2-b_3G_3\]
has the leading term defined by the curvature tensor $\mathcal R$ (\ref{ham-gen}, \ref{ham-13bb})
\[\begin{array}{rcl}
G_4&=&-\frac14\sum_{\alpha,\beta,\gamma,\delta} \mathcal R_{-\alpha,\beta,\gamma,-\delta}p^\alpha p^\beta p^\gamma p^\delta+\cdots\\
\\
&=&\frac{1}{2}(p_1^2 + p_2^2 + p_3^2)^2 + \frac{1}{2}(p_4^2 + p_5^2 + p_6^2)^2 + (p_1p_4 + p_2p_5 + p_3p_6)^2+\cdots.
\end{array}
\]
Summing up, there are only $m+n-1=4$ independent quadratic integrals of motion among $f_1,f_2,f_3$ and $F_1,F_2,F_3,F_4$ in $\mathbb R^6$. The corresponding Killing tensors of valency two have non-trivial Haantjes torsion.

\subsection{Euclidean space $ \mathbb R^9$, case $m=n=3$}
The $6\times 6$ Lax matrix (\ref{lax-a}) is
\bq\label{lax-6}
L(\lambda)=\left(
\begin{matrix}
L_{11}&L_{12}\\
L_{21}&L_{22}
\end{matrix}\right)
\eq
where
\[\begin{array}{rcl}
L_{11}&=&\left(\begin{smallmatrix}-2 \lambda^2 + q_1^2 + q_2^2 + q_3^2 + a_1& q_1 q_4 + q_2 q_5 + q_3 q_6& q_1 q_7 + q_2 q_8 + q_3 q_9\\
q_1 q_4 + q_2 q_5 + q_3 q_6& -2 \lambda^2 + q_4^2 + q_5^2 + q_6^2 + a_2& q_4 q_7 + q_5 q_8 + q_6 q_9\\
q_1 q_7 + q_2 q_8 + q_3 q_9& q_4 q_7 + q_5 q_8 + q_6 q_9& -2 \lambda^2 + q_7^2 + q_8^2 + q_9^2 + a_3& \\
\end{smallmatrix}\right)\,,
\\
\\
L_{22}&=& \left(\begin{smallmatrix}
 \phantom{-}2 \lambda^2 - q_1^2 - q_4^2 - q_7^2 + b_1& -q_1 q_2 - q_4 q_5 - q_7 q_8& -q_1 q_3 - q_4 q_6 - q_7 q_9\\
 -q_1 q_2 - q_4 q_5 - q_7 q_8& \phantom{-}2 \lambda^2 - q_2^2 - q_5^2 - q_8^2 + b_2& -q_2 q_3 - q_5 q_6 - q_8 q_9\\
 -q_1 q_3 - q_4 q_6 - q_7 q_9& -q_2 q_3 - q_5 q_6 - q_8 q_9& \phantom{-} 2 \lambda^2 - q_3^2 - q_6^2 - q_9^2 + b_3\\
 \end{smallmatrix}\right)\,,
\\
\\
L_{12}&=& \left(
\begin{smallmatrix} p_1 - 2\mathrm i \lambda q_1& p_2 - 2\mathrm i \lambda q_2& p_3 - 2\mathrm i \lambda q_3\\ \\
 p_4 - 2\mathrm i \lambda q_4& p_5 - 2\mathrm i \lambda q_5& p_6 - 2\mathrm i \lambda q_6\\ \\
p_7 - 2\mathrm i \lambda q_7& p_8 - 2\mathrm i \lambda q_8& p_9 - 2\mathrm i \lambda q_9\\
\end{smallmatrix}\right)\,,\qquad
L_{21}=\left(\begin{smallmatrix}
p_1 + 2\mathrm i \lambda q_1& p_4 + 2\mathrm i \lambda q_4& p_7 + 2\mathrm i \lambda q_7\\ \\
p_2 + 2\mathrm i \lambda q_2& p_5 + 2\mathrm i \lambda q_5& p_8 + 2\mathrm i \lambda q_8\\ \\
p_3 + 2\mathrm i \lambda q_3& p_6 + 2\mathrm i \lambda q_6& p_9 + 2\mathrm i \lambda q_9\\
\end{smallmatrix}\right)\,.
\end{array}
\]
Hamiltonian $H$ (\ref{ham-gena}) is given by
\begin{align*}
H=&\frac12 \sum_{i=1}^9 p_i^2+\frac{(q_1^2 + q_2^2 + q_3^2)^2}{2} +\frac{(q_4^2 + q_5^2 + q_6^2)^2}{2} +\frac{(q_7^2 + q_8^2 + q_9^2)^2}{2}
\nn\\
\nn\\
+& (q_1q_4 + q_2q_5 + q_3q_6)^2+ (q_1q_7 + q_2q_8 + q_3q_9)^2 + (q_4q_7 + q_5q_8 + q_6q_9)^2\nn\\
\nn\\
-&\frac{q_1^2+ q_4^2+ q_7^2}{2}b_1 -\frac{q_2^2+ q_5^2+ q_8^2}{2}b_2 -\frac{q_3^2+ q_6^2+ q_9^2}{2}b_3\nn\\
\nn\\
+&\frac{q_1^2 + q_2^2+ q_3^2}{2}a_1 +\frac{q_4^2 + q_5^2 + q_6^2}{2}a_2 + \frac{q_7^2 + q_8^2 + q_9^2}{2}a_3\,.\nn
\end{align*}
\vskip0.2truecm
\par\noindent\textbf{First set of integrals of motion}
\par\noindent
Residues of the function $\Delta(z,\lambda)$ (\ref{delta1})
\[
\Delta(z,\lambda)=\frac{\det\Bigl(zI-L(\lambda)\Bigr)}{(z-a_1+2\lambda^2)(z-a_2+2\lambda^2)(z-a_3+2\lambda^2)}
\]
are equal to
\begin{align*}
\left.\mbox{Res}\right|_{z=a_i+2\lambda^2}\, \Delta(z,\lambda)=16\lambda^4 f_i+\lambda^2 g_i+s_i\,,\qquad i=1,2,3.
\\
\\
\left.\mbox{Res}\right|_{z=\infty}\, \Delta(z,\lambda)=32H\lambda^4 -(g_1+g_2+g_3)\lambda^2-(s_1+s_2+s_3)\,.
\end{align*}
where second-order polynomials in momenta are equal to
\begin{align*}
f_1=&\frac{M_{12}^2}{a_1 - a_2} +\frac{M_{13}^2}{a_1 - a_3}-p_1^2 - p _{2}^2 - p_{3}^2
-(2{q_2}^{2}+2{q_3}^{2}+{q_4}^{2}+{q_7}^{2}+a_1-b_1){q_1}^{2}
\nn\\
\nn\\
 -& (2 {q_3}^{2}-{q_5}^{2}-{q_8}^{2}-{a_1}+{b_2}){q_2}^{2} -({q_3}^{2}+{q_6}^{2}+{q_9}^{2}+{a_1}-{b_3}) {q_3}^{2}
 \nn\\
 \nn\\
 - & 2{q_2}{q_3} ({q_5}{q_6}+{q_8} {q_9})- 2 q_1 q_2 ( q_4 q_5 + q_7 q_8) - 2 q_1 q_3 (q_4 q_6+q_7 q_9)- {q_1}^{4}-{q_2}^{4}\,,
\nn\\
\nn\\
f_2=&\frac{M_{21}^2}{a_2 - a_1} +\frac{M_{23}^2}{a_2 - a_3}-p_4^2 - p_5^2 - p_6^2
-({q_1}^{2}+2{q_5}^{2}+2{q_6}^{2}+{q_7}^{2}+{a_2}-{b_1}){q_4}^{2}
\nn\\
\nn\\
 -& ({q_2}^{2}+2 {q_6}^{2}+{q_8}^{2}+{a_2}-{b_2}){q_5}^{2} - (q_3^2 + q_6^2 + q_9^2 + a_2 - b_3) q_6^2
\nn\\
\nn\\
-& q_5 q_4 (2 q_1 q_2 + 2 q_7 q_8) - 2 q_6 q_4(q_1 q_3 + q_7 q_9) - 2 q_5 q_6 (q_2 q_3 + q_8 q_9) - q_4^4-q_5^4
\nn
\end{align*}
and
\begin{align*}
f_3=&\frac{M_{31}^2}{a_3 - a_1} +\frac{M_{32}^2}{a_3 - a_2}-p_7^2 - p_8^2 - p_9^2
-(q_1^2 + q_4^2 + 2 q_8^2 + 2 q_9^2 + a_3 - b_1) q_7^2
\nn\\
\nn\\
-& (q_2^2 + q_5^2 + 2 q_9^2 + a_3 - b_2) q_8^2
 - (q_3^2 + q_6^2 + q_9^2 + a_3 - b_3) q_9^2
 \nn\\
 \nn\\
 -& 2q_7 q_8(q_1q_2 + q_4q_5) - 2q_7q_9(q_1q_3 + q_4q_6) - 2q_8q_9(q_2q_3 + q_5q_6) - q_7^4 - q_8^4\,,
 \nn
\end{align*}
Here $M_{ij}$ are given by
\begin{align}
M_{12}=& J_{14}+J_{25}+J_{36}=(q_1p_4-p_1q_4) + (q_2p_5-p_2q_5)+ (q_3p_6-p_3q_6)\,,\nn\\
\nn\\
M_{13}=&J_{17}+J_{28}+J_{39}=(q_1p_7-p_1q_7)+(q_2p_8-p_2q_8)+ (q_3p_9-p_3q_9)\,,\label{fm-so3-3}\\
\nn\\
M_{23}=&J_{47} +J_{58}+J_{69}=(q_4p_7-p_4q_7)+(q_5p_8-p_5q_8)+ (q_6p_9-p_6q_9)\,.\nn
\end{align}
The following combination of integrals of motion is also a quadratic polynomial in momenta
\[
f_4= \frac{g_1 + g_2 + g_2}{4} + 2a_1f_1 + 2a_2f_2+2a_3 f_3\,,
\]
which has the form
\[
f_4=-\left(\sum_{j=1}^{n} b_j\right)\left(\sum_{i=1}^{nm} p_i^2\right)
+\sum_{j=1}^n b_{j}\left(\sum_{i=0}^{m-1} p_{j+im}^2\right)
+N_{12}^2+N_{23}^2+N_{31}^2+u_4(q)\,.
\]
Here $N_{ij}$:
\begin{align}
N_{12}=& J_{12}+J_{45}+J_{78}=(q_1p_2-p_1q_2) + (q_4p_5-p_4q_5)+ (q_7p_8-p_7q_8)\,,\nn\\
\nn\\
N_{13}=&J_{13}+J_{46}+J_{79}=(q_1p_3-p_1q_3)+(q_4p_6-p_4q_6)+ (q_7p_9-p_7q_9)\,,\label{fn-so3-3}\\
\nn\\
N_{23}=&J_{23} +J_{56}+J_{89}=(q_2p_3-p_2q_3)+(q_5p_6-p_5q_6)+ (q_8p_9-p_8q_9)\,.\nn
\end{align}
Functions $M_{ij}$ (\ref{fm-so3-3}) and $N_{ij}$ (\ref{fn-so3-3}) are associated with two realizations of $so^*(3)$ by using independent triple rotations in $ \mathbb R^9$. The Lie-Poisson brackets are
\[
\{M_{12}, M_{13}\} = M_{23}\,,\qquad \{M_{13}, M_{23}\}=M_{12}\,,\qquad
\{M_{23}, M_{12}\}=M_{13}\,.
\]
and
\[
\{N_{12}, N_{13}\} = N_{23}\,,\qquad \{N_{13}, N_{23}\}=N_{12}\,,\qquad
\{N_{23}, N_{12}\}=N_{13}\,,
\]
so that
\[
\{N_{ij}, M_{kl}\}=0\,.
\]
Leading terms in polynomials of six order in momenta are
\[
s_i=\frac{1}{(a_i-a_j)(a_i-a_k)}\,\bigl(p_1p_5p_9+ p_2p_6p_7+ p_3p_4p_8 - p_1p_6p_8 - p_2p_4p_9 - p_3p_5p_7
\bigr)^2+\cdots\,,
\]
and, therefore, the sum of these polynomials is a polynomial of the fourth order in momenta which is independent of $g_1,g_2$ and $g_3$.
\vskip0.2truecm
\par\noindent\textbf{Second set of the integrals of motion}
\par\noindent
Residues of the function $\Delta(z,\lambda)$ (\ref{delta2})
\[
\Delta(z,\lambda)=\frac{\det\Bigl(zI-L(\lambda)\Bigr)}{(z-b_1-2\lambda^2)(z-b_2-2\lambda^2)(z-b_3-2\lambda^2)}
\]
 are equal to
\begin{align*}
\left.\mbox{Res}\right|_{z=b_i+2\lambda^2}\, \Delta(z,\lambda)=16\lambda^4 F_i+\lambda^2 G_i+S_i\,,\qquad i=1,2,3.
\\
\\
\left.\mbox{Res}\right|_{z=\infty}\, \Delta(z,\lambda)=32H\lambda^4 -(G_1+G_2+G_3)\lambda^2-(S_1+S_2+S_3)\,.
\end{align*}
Second-order polynomials in momenta have the following form
\begin{align*}
F_1&=-\frac{N_{12}^2}{b_1 - b_2} -\frac{N_{13}^2}{b_1 - b_3}-p_1^2 - p _{4}^2 - p_{7}^2
-(q_1^2 + q_2^2 + q_3^2 + q_4^2 + q_7^2 +a_1- b_1 )q_1^2
\nn\\
\nn\\
& - (q_1^2 + q_4^2 + q_5^2 + q_6^2 + q_7^2+a_2 - b_1)q_4^2
 - (q_1^2 + q_4^2 + q_7^2 + q_8^2 + q_9^2+a_3 - b_1)q_7^2
 \nn\\
 \nn\\
 &- 2q_1q_4(q_2q_5 + q_3q_6)- 2q_1q_7(q_2q_8 + q_3q_9)- 2q_4q_7(q_5q_8 + q_6q_9)\,,
\nn\\
\nn\\
F_2&=-\frac{N_{21}^2}{b_2 - b_1} -\frac{N_{23}^2}{b_2 - b_3}-p_2^2 - p_5^2 - p_8^2
-(q_1^2 + q_2^2 + q_3^2 + q_5^2 + q_8^2+a_1 - b_2)q_2^2
\nn\\
\nn\\
& - (q_2^2 + q_4^2 + q_5^2 + q_6^2 + q_8^2+a_2-b_2)q_5^2 - (q_2^2 + q_5^2 + q_7^2 + q_8^2 + q_9^2+a_3 - b_2)q_8^2
\nn\\
\nn\\
&- 2q_2q_5(q_1q_4 + q_3q_6) - 2q_2q_8(q_1q_7 + q_3q_9) - 2q_5q_8(q_4q_7 + q_6q_9)
\nn
\end{align*}
and
\begin{align*}
F_3&=-\frac{N_{31}^2}{b_3 - b_1} -\frac{N_{32}^2}{b_3 - b_2}-p_3^2 - p_6^2 - p_9^2
-(q_1^2 + q_2^2 + q_3^2 + q_6^2 + q_9^2 +a_1- b_3)q_3^2
\nn\\
\nn\\
&- (q_3^2 + q_4^2 + q_5^2 + q_6^2 + q_9^2+a_2 - b_3)q_6^2
 - (q_3^2 + q_6^2 + q_7^2 + q_8^2 + q_9^2+a_3 - b_3)q_9^2
 \nn\\
 \nn\\
 & - 2q_3q_6(q_1q_4 + q_2q_5) - 2q_3q_9(q_1q_7 + q_2q_8) - 2q_6q_9(q_4q_7 + q_5q_8)\,,
 \nn
\end{align*}
Functions $M_{ij}, N_{kl}$ are given by \ref{fm-so3-3},\ref{fn-so3-3}.

The following combination of integrals of motion is also second order polynomial in momenta
\[
F_4=\frac18(G_1 + G_2 + G_3) - b_1F_1 - b_2F_2 - b_3F_3
\]
which is independent on $F_1,F_2$ and $F_3$. It has the form
\ben
\nn
F_4=\frac12\left(\sum_{j=1}^{m} a_j\right)\left(\sum_{i=1}^n p_i^2\right)
-\frac12\sum_{j=0}^{m-1}a_{j+1}\left(\sum_{i=1}^n p_{jn+i}^2\right)
+\frac{M_{12}^2}{2}+\frac{M_{13}^2}{2}+\frac{M_{23}^2}{2}+U_4(q)\,.
\nn
\en
Leading terms in polynomials of six order in momenta are
\[
S_i=\frac{1}{(b_i-b_j)(b_i-b_k)}\,\bigl(p_1p_5p_9+ p_2p_6p_7+ p_3p_4p_8 - p_1p_6p_8 - p_2p_4p_9 - p_3p_5p_7
\bigr)^2+\cdots\,,
\]
and, therefore, the sum of these polynomials is a polynomial of the fourth order in momenta
\[G_4=S_1+S_2+S_3\,,\]
which is independent on $g_1,g_2$ and $g_3$.

Summing up, we have integrals of motion of second, fourth-order and sixth order in momenta. Among them, there are five independent quadratic integrals of motion because
\[f_1+f_2+f_3=2H=F_1+F_2+F_3\]

\section{Symmetric space of C.I type}
\setcounter{equation}{0}

The compact group $Sp(n)$ of $2n \times 2n$ matrices which are both symplectic and unitary is associated with the root space $C_n$. Because
\[
\frac{Sp(n)}{U(n)}\subset \frac{SU(2n)}{S\left(U(n)\times U(n)\right)}
\]
we can get the Lax matrices starting with Lax matrices (\ref{lax-a}). Roughly speaking we have to make $n\times n$ matrices $Q$ and $P$ symmetric,
divide off-diagonal entries of $P$ by two and impose suitable restrictions on parameters $a_i$ and $b_i$.

Below we present these Lax matrices at $n=2$ and $n=3$ and discuss the corresponding quadratic integrals of motion.

\subsection{Euclidean space $\mathbb R^3$, case $n=2$}
The $4\times 4$ Lax matrix is equal to
\bq\label{lax-13c}
L(\lambda)=\left(
 \begin{smallmatrix}
 -2\lambda^2+ q_1^2 + q_2^2 +a_1& q_1q_2 + q_2q_3 & p_1 - 2\mathrm i \lambda q_1 & \frac{p_2}{2} - 2\mathrm i\lambda q_2 \\
 \\
 q_1q_2 + q_2q_3 &-2\lambda^2+q_2^2 + q_3^2+a_2 &\frac{p_2}{2} - 2\mathrm i\lambda q_2 & p_3 - 2\mathrm i \lambda q_3 \\
 \\
 p_1 + 2\mathrm i \lambda q_1 & \frac{p_2}{2} + 2\mathrm i\lambda q_2 & 2\lambda^2 - q_1^2 - q_2^2+b_1 & -q_1q_2 - q_2q_3 \\
 \\
 \frac{p_2}{2} +2\mathrm i\lambda q_2 & p_3 + 2\mathrm i \lambda q_3 & -q_1q_2 - q_2q_3 & 2\lambda^2 - q_2^2 - q_3^2 +b_2\\
 \end{smallmatrix}
 \right)\,,
\eq
where
\bq\label{cond-ab2} a_i=-b_i \,.
\eq
The Hamiltonian is given by
\bq\label{ham-13c}
\begin{array}{rcl}
H=T+V&=&\dfrac{p_1^2}{2} +\dfrac{p_2^2}{4} +\dfrac{p_3^2}{2}+\dfrac{(q_1^2 + 2q_2^2 + q_3^2)^2}{2}
\\
\\
&-& (q_1q_3 - q_2^2)^2-b_1(q_1^2 +q_2^2)-b_2(q_2^2 + q_3^2)\,.
\end{array}
\eq
It coincides with the (13c) case from the paper \cite{f86}.

After canonical change of variables $p_2\to \sqrt{2} p_2$ and $q_2\to q_2/\sqrt{2}$
we obtain standard metric $\mathrm g=diag(1,1,1)$ in Euclidean space and integrable three-dimensional quartic potential at $b_i=a_i=0$
\bq\label{c-pot}
V=\frac12(q_1^2 + q_2^2 + q_3^2)^2 -\frac{ (2q_1q_3 - q_2^2)^2}{4}
\eq
This potential is missing in the classification based on the Ziglin and Yoshida methods \cite{d86}, since authors studied only potentials in the following form
\[
\tilde{V}=q_1^4+aq_1^2q_2^2+bq_1^2q_3^2+cq_2^4+dq_2^2q_3^2+eq_3^4\,,
\]
whereas (\ref{c-pot}) involves linear in $q_1$ and $q_3$ term $q_1q_3q_2^2$.

Residues of the functions
\[
\Delta(z,\lambda)=\frac{\det\Bigl(Iz-L(\lambda)\Bigr)}{(z+2\lambda^2 -a_1)(z+2\lambda^2 -a_2)}
\qquad\mbox{and}\qquad
\Delta(z,\lambda)=\frac{\det\Bigl(Iz-L(\lambda)\Bigr)}{(z-2\lambda^2 -b_1)(z-2\lambda^2 -b_2)}
\]
coincide for each other up to the sign and replacement $a_1-a_2=-(b_1-b_2)$ which corresponds to (\ref{cond-ab2}).

Let us consider residues
\begin{align*}
&\left.\mbox{Res}\right|_{z=b_i+2\lambda^2}\, \Delta(z,\lambda)=-4\lambda^2 F_i+G_i\,,\qquad i=1,2.
\\
\\
&\left.\mbox{Res}\right|_{z=\infty}\, \Delta(z,\lambda)=8\lambda^2 H -(G_1+G_2)\,.
\end{align*}
Because
\[
 F_1 + F_2 - 2H=0\qquad\mbox{and}\qquad G_1+G_2+2(b_2-a_1)H=0\,.
\]
there are two polynomials of the second order in momenta $F_{1,2}$ and only one polynomial of the fourth order in momenta $G_1$ or $G_2$ which are independent of each other.

In \cite{f86} authors argue that three integrals of motion $F_1$, $F_2$ and $G_1+G_2$ are quadratic polynomials in momenta, thus suggesting the existence of the point transformation to new variables in which equations of motion (\ref{eqm}) can be separated. Unfortunately, the authors did not notice that these integrals of motion are functionally dependent; therefore, their statement is incorrect.

Let us present these integrals explicitly

\begin{align*}
F_1=p_1^2 + \frac{p_2^2}{4} + \frac{M_{12}^2}{b_1-b_2}+(q_1^2 + 2q_2^2 + a_1 - b_1)q_1^2 + (q_1^2 + q_2^2 + q_3^2 + 2q_1q_3 + a_1 - b_2)q_2^2\,,\\
\\
F_2=p_3^2+\frac{p_2^2}{4}+\frac{M_{12}^2}{b_2-b_1} + (2q_2^2 + q_3^2 + a_2 - b_2)q_3^2 + (q_1^2 + q_2^2 + q_3^2+ 2q_1q_3 + a_2 - b_1)q_2^2\,.
\end{align*}
where
\[M_{12}= \frac{1}{2}(q_1p_2-2q_2p_1 -q_3p_2+ 2q_2p_3)\,.
\]
At $b_1=b_2$ we have a linear integral of motion $M_{12}$ associated with a double rotation. After reduction by the corresponding Noether's symmetry, we obtain new quadratic-linear Hamiltonian $H$ commuting with the quartic invariant $G_{1,2}$ in $T^*\mathbb R^2$

For Euclidean space $\mathbb R^3$ generic solution $K$ (\ref{kill2-gen}) of the Killing equation (\ref{kill-eq}) depends on 20 parameters. Using modern computer software we can directly prove that there are only two independent solutions to the equation
\[d(KdV)=0\]
associated with the integrals of motion $F_{1,2}$.

Moreover, by substituting into this equation (\ref{dkdv}) the Killing tensors associated to $F_{1,2}$ and the unknown function $V(q_1,q_2,q_3)$ we obtain a more general potential. Indeed, consider quadratic integrals of motion 
\[H=p_1^2+p_2^2 +p_3^2 +V(q)\]
and
\begin{align*}
F_1&=2p_1^2+p_2^2  + \frac{(q_1p_2-p_1q_2+q_2p_3 - p_2q_3 )^2}{b_1-b_2}+U_1(q)\,,\\ \\
F_2&=2p_3^2+p_2^2  + \frac{(q_1p_2-p_1q_2+q_2p_3 - p_2q_3 )^2}{b_2-b_1}+U_2(q)\,.
\end{align*}
The general solution of the equation (\ref{dkdv}) depends on five arbitrary parameters
\begin{align}
V&=c_1\left(q_1^4 + 2q_1^2q_2^2 + 2q_1q_2^2q_3 + \frac{q_2^4}{2} + 2q_2^2q_3^2 + q_3^4 -2 (q_1^2 + q_2^2)b_1 + 2(q_1^2 + q_2^2)b_2\right)
\nn\\ \nn\\
&+c_2\left(2q_1^3 + 3q_2^2(q_1+q_3) + 2q_3^3 - 2b_1q_1 + 2b_2q_1\right)
+c_3\left(q_1^2+q_2^2+q_3^2\right)\label{v-cub}\\ \nn\\
&+c_4(q_1+q_3)+\frac{c_5}{q_2^2}\,,\qquad c_i\in\mathbb R\,.\nn
\end{align}
Because we know leading part of the third integral of motion (\ref{G4}),   we can find an explicit generic expression for this integral of motion depending on $c_1,\ldots,c_5$ and thus prove the integrability of this Hamiltonian system.

\subsection{Euclidean space $\mathbb R^6$, case $n=3$}
The $6\times 6$ Lax matrix (\ref{lax-6}) generates three quadratic invariants $F_i$, three quartic $G_i$ and three sextic invariants $S_i$. After reduction, we have to get only six independent invariants.

The Lax matrix (\ref{lax-6}) after reduction looks like
\[
L(\lambda)=\left(
\begin{matrix}
\hat L_{11}&\hat L_{12}\\
\hat L_{21}& \hat L_{22}
\end{matrix}\right)
\]
where
\[\begin{array}{rcl}
\hat L_{11}&=&\left(\begin{smallmatrix}-2 \lambda^2 + q_1^2 + q_2^2 + q_3^2 + a_1& q_1q_2 + q_2q_4 + q_3q_5& q_1q_3 + q_2q_5 + q_3q_6 \\
 q_1q_2 + q_2q_4 + q_3q_5& -2\lambda^2 + q_2^2 + q_4^2 + q_5^2 + b_1 - b_2 + a_1& q_2q_3 + q_4q_5 + q_5q_6 \\
q_1q_3 + q_2q_5 + q_3q_6& q_2q_3 + q_4q_5 + q_5q_6&-2\lambda^2 + q_3^2 + q_5^2 + q_6^2 + b_1 - b_3 + a_1 \\
\end{smallmatrix}\right)\,,
\\
\\
\hat L_{22}&=& \left(\begin{smallmatrix}
 \phantom{-}2\lambda^2 - q_1^2 - q_2^2 - q_3^2 + b_1&-q_1q_2 - q_2q_4 - q_3q_5 &-q_1q_3 - q_2q_5 - q_3q_6 \\
 -q_1q_2 - q_2q_4 - q_3q_5& \phantom{-}2\lambda^2 - q_2^2 - q_4^2 - q_5^2 + b_2&-q_2q_3 - q_4q_5 - q_5q_6 \\
-q_1q_3 - q_2q_5 - q_3q_6&-q_2q_3 - q_4q_5 - q_5q_6 & \phantom{-} 2\lambda^2 - q_3^2 - q_5^2 - q_6^2 + b_3\\
 \end{smallmatrix}\right)\,,
\\
\\
\hat L_{12}&=& \left(
\begin{smallmatrix} p_1 - 2\mathrm i \lambda q_1& \frac{p_2}{2} - 2\mathrm i \lambda q_2& \frac{p_3}{2} - 2\mathrm i \lambda q_3\\ \\
 \frac{p_2}2 - 2\mathrm i \lambda q_2& p_4 - 2\mathrm i \lambda q_4& \frac{p_5}{2} - 2\mathrm i \lambda q_5\\ \\
\frac{p_3}{2} - 2\mathrm i \lambda q_3& \frac{p_5}{2} - 2\mathrm i \lambda q_5& p_6 - 2\mathrm i \lambda q_6\\
\end{smallmatrix}\right)\,,\qquad
\hat L_{21}=\left(
\begin{smallmatrix} p_1 + 2\mathrm i \lambda q_1& \frac{p_2}{2}+ 2\mathrm i \lambda q_2& \frac{p_3}{2} + 2\mathrm i \lambda q_3\\ \\
 \frac{p_2}2 + 2\mathrm i \lambda q_2& p_4 + 2\mathrm i \lambda q_4& \frac{p_5}{2} + 2\mathrm i \lambda q_5\\ \\
\frac{p_3}{2} + 2\mathrm i \lambda q_3& \frac{p_5}{2} + 2\mathrm i \lambda q_5& p_6 +2\mathrm i \lambda q_6\\
\end{smallmatrix}\right)\,.
\end{array}
\]
Here we impose the following restrictions on arbitrary parameters in (\ref{lax-6})
\[a_i=-b_i\,.\]
Calculating integrals of motion
 using three residues of the function
\[
\Delta=\frac{\det\Bigl(Iz-L(\lambda)\Bigr)}{(z-2\lambda^2 - b_1)(z-2\lambda^2 - b_2)(z-2\lambda^2 - b_3)}
\]
at $z=b_i+2\lambda^2$
\[
\left.\mbox{Res}\right|_{z=b_i+2\lambda^2}\, \Delta(z,\lambda)=-16\lambda^4 F_i+\lambda^2 G_i+ S_i\]
we obtain the following quadratic integrals of motion
\begin{align*}
F_1=&\frac{M_{12}^2}{b_1 - b_2} +\frac{M_{13}^2}{b_1-b_3}+T_1+V_1\,,\qquad T_1=p_1^2 + \frac{p_2^2}{4} + \frac{p_3^2}{4}\,,\\ \\
F_2=&\frac{M_{21}^2}{b_2 - b_1} +\frac{M_{23}^2}{b_2-b_3}+T_2+V_2\,,\qquad T_2=\frac{p_2^2}{4}+ p_4^2 +\frac{p_5^2}{4}\,,\\ \\
F_3=&\frac{M_{31}^2}{b_3 - b_1} +\frac{M_{32}^2}{b_3-b_2}+T_3+V_3\,,\qquad T_3=\frac{p_3^2}{4}+ \frac{p_5^2}{4} + p_6^2\,,\\
\end{align*}
where functions associated with the triple rotations are equal to
\begin{align*}
M_{12}&=-M_{21}= \frac12 (q_1p_2- 2p_1q_2 +2q_2p_4- p_2q_4+q_3p_5 - p_3q_5)\,,\\
M_{13}&=-M_{31}=\frac12 (q_1p_3 -2p_1q_3 +q_2p_5- p_2q_5 +2q_3p_6- p_3q_6)\,,\\
M_{23}&=-M_{32}=\frac12 (q_2p_3-p_2q_3+q_4p_5 - 2p_4q_5 +2q_5p_6 - p_5q_6)\,.
\end{align*}
For brevity, we omit explicit expressions for the potentials $V_k$.

Residue at infinity gives rise to a relation between quadratic integrals
\[F_1+F_2+F_3-2H=0\]
and relations between other integrals of motion
\[
\frac{1}{4}\,(G_1+G_2+G_3) + b_1F_1 + b_2F_2 + b_3F_3 + 2(b_1 + b_2 + b_3)H=0
\]
and
\ben
S_1 + S_2 + S_3 &-&\frac14(b_1G_1 +b_2 G_2 +b_3 G_3)\nn\\
\nn\\
&-& (b_1^2 - b_2b_3)F_1 - (b_2^2-b_1b_3)F_2 -(b_3^2-b_1b_2)F_3=0\,.\nn
\en
From nine dependent integrals of motion $F_i, G_i$ and $S_i$ we have to choose six independent, for instance, we can take three quadratic integrals of motion, two integrals of motion of fourth order and one integral of sixth order in momenta.

\section{Symmetric space of D.III type}
\setcounter{equation}{0}
This is another reduction of the A.III case
\[
\frac{SO(2n)}{U(n)}\subset \frac{SU(2n)}{S\left(U(n)\times U(n)\right)}
\]
associated with the root space $D_n$. In this case we have to take Lax matrices (\ref{lax-a}), make $n\times n$ matrices $Q$ and $P$ antisymmetric,
and impose suitable restrictions on parameters $a_i$ and $b_i$.

Isomorphism $D_3\cong A_3$ yields a correspondence
\[
\frac{SO(6)}{U(3)}\cong\frac{SU(4)}{S(U(1)\times U(3))}
\]
so that we have the well-known Garnier system in $\mathbb R^3$ and the corresponding Hamilton-Jacobi equation $H=E$ is separable in the elliptic coordinates.

Following \cite{f83} we restrict ourselves by calculation of the quadratic integrals of motion for the $D_4$ case.

\subsection{Euclidean space $\mathbb R^6$, case $n=4$}
The $8\times 8$ Lax matrix (\ref{lax-a}) after reduction has the following form
\[
L(\lambda)=\left(
\begin{matrix}
\bar L_{11}&\bar L_{12}\\
\bar L_{21}& \bar L_{22}
\end{matrix}\right)\,.
\]
There are two symmetric matrices
\[
\bar{L}_{11}= \left( \begin{smallmatrix}
 q_1^2 + q_2^2 + q_3^2 + a_1 -2 \lambda^2 & q_2 q_4 + q_3 q_5 & -q_1 q_4 + q_3 q_6 & -q_1 q_5 - q_2 q_6 \\
 q_2 q_4 + q_3 q_5 & q_1^2 + q_4^2 + q_5^2 + a_2 -2 \lambda^2 & q_1 q_2 + q_5 q_6 & q_1 q_3 - q_4 q_6 \\
 -q_1 q_4 + q_3 q_6 & q_1 q_2 + q_5 q_6 & q_2^2 + q_4^2 + q_6^2 + a_3 -2 \lambda^2 & q_2 q_3 + q_4 q_5 \\
 -q_1 q_5 - q_2 q_6 & q_1 q_3 - q_4 q_6 & q_2 q_3 + q_4 q_5 & q_3^2 + q_5^2 + q_6^2 + a_4 -2 \lambda^2 \\
 \end{smallmatrix}\right)\,,
\]
\[
\bar{L}_{22}=\left(
 \begin{smallmatrix}
 2 \lambda^2 - q_1^2 - q_2^2 - q_3^2 + b_1 & -q_2 q_4 - q_3 q_5 & q_1 q_4 - q_3 q_6 & q_1 q_5 + q_2 q_6 \\
 -q_2 q_4 - q_3 q_5 & 2 \lambda^2 - q_1^2 - q_4^2 - q_5^2 + b_2 & -q_1 q_2 - q_5 q_6 & -q_1 q_3 + q_4 q_6 \\
 q_1 q_4 - q_3 q_6 & -q_1 q_2 - q_5 q_6 & 2 \lambda^2 - q_2^2 - q_4^2 - q_6^2 + b_3 & -q_2 q_3 - q_4 q_5 \\
 q_1 q_5 + q_2 q_6 & -q_1 q_3 + q_4 q_6 & -q_2 q_3 - q_4 q_5 & 2 \lambda^2 - q_3^2 - q_5^2 - q_6^2 + b_4 \\
 \end{smallmatrix}
 \right)
\]
and two antisymmetric matrices
\[
\bar{L}_{12}=\left(
 \begin{smallmatrix}
 0 & p_1 - 2\mathrm i \lambda q_1 & p_2 - 2\mathrm i \lambda q_2 & p_3 - 2\mathrm i \lambda q_3 \\
 -p_1 + 2\mathrm i \lambda q_1 & 0 & p_4 - 2\mathrm i \lambda q_4 & p_5 - 2\mathrm i \lambda q_5 \\
 -p_2 + 2\mathrm i \lambda q_2 & -p_4 + 2\mathrm i \lambda q_4 & 0 & p_6 - 2\mathrm i \lambda q_6 \\
 -p_3 + 2\mathrm i \lambda q_3 & -p_5 + 2\mathrm i \lambda q_5 & -p_6 + 2\mathrm i \lambda q_6 & 0 \\
 \end{smallmatrix}
 \right)\,,
\]
\[
\bar{L}_{21}=\left(
 \begin{smallmatrix}
 0 & -p_1 - 2\mathrm i \lambda q_1 & -p_2 - 2\mathrm i \lambda q_2 & -p_3 - 2\mathrm i \lambda q_3 \\
 p_1 + 2\mathrm i \lambda q_1 & 0 & -p_4 - 2\mathrm i \lambda q_4 & -p_5 - 2\mathrm i \lambda q_5 \\
 p_2 + 2\mathrm i \lambda q_2 & p_4 + 2\mathrm i \lambda q_4 & 0 & -p_6 - 2\mathrm i \lambda q_6 \\
 p_3 + 2\mathrm i \lambda q_3 & p_5 + 2\mathrm i \lambda q_5 & p_6 + 2\mathrm i \lambda q_6 & 0 \\
 \end{smallmatrix}
 \right)\,.
\]
The parameters must satisfy the following constraints
\[a_2-a_1= b_1 - b_2\,,\qquad a_3-a_1= b_1 - b_3\,,\qquad a_4-a_1= b_1 - b_4\,.
\]
Four residues of the function
\[
\Delta=\frac{\det\Bigl(Iz-L(\lambda)\Bigr)}{(z-2\lambda^2 - b_1)(z-2\lambda^2 - b_2)(z-2\lambda^2 - b_3)(z-2\lambda^2 - b_4)}
\]
at $z=b_i+2\lambda^2$ are polynomials of sixth order in momenta
\[
\left.\mbox{Res}\right|_{z=b_i+2\lambda^2}\, \Delta(z,\lambda)=-64\lambda^6 F_i+\lambda^4 G_i+\lambda^2 S_i+W_i\,,
\]
where $F_i$, $G_i$, $S_i$ and $W_i$ are the second, fourth, sixth and eighth-order polynomials in momenta. As a result, we have 16 dependent integrals of motion and residue at infinity yields various relations between these polynomials, for instance
\[F_1+F_2+F_3+F_4-2H=0\,.\]
We show only the leading part of the quadratic integrals of motion and omit explicit expressions for the potentials $V_k$
\begin{align*}
F_1&=\frac{M_{12}^2}{b_1-b_2}+\frac{M_{13}^2}{b_1-b_2}+\frac{M_{14}^2}{b_1-b_4}+T_1+V_1\,,\quad
F_2=\frac{M_{21}^2}{b_2-b_1}+\frac{M_{23}^2}{b_2-b_3}+\frac{M_{24}^2}{b_2-b_4}+T_2+V_2\,,\\ \\
F_3&=\frac{M_{31}^2}{b_3-b_1}+\frac{M_{32}^2}{b_3-b_2}+\frac{M_{34}^2}{b_3-b_4}+T_3+V_3\,,\quad
F_4=\frac{M_{41}^2}{b_4-b_1}+\frac{M_{42}^2}{b_4-b_2}+\frac{M_{43}^2}{b_4-b_3}+T_4+V_4\,.
\end{align*}
where functions
\begin{align*}
M_{12}=&(q_2p_4-p_2q_4)+(q_3p_5-p_3q_5)\,,\qquad
M_{13}=(q_1p_4-p_1q_4)+ (q_6p_3-p_6q_3)\,,\\
M_{14}=&(q_1p_5-p_1q_5)+(q_2p_6-p_2q_6)\,,\qquad
M_{23}=(q_1p_2- p_1q_2) + (q_5p_6-p_5q_6)\,,\\
M_{24}=&(q_1p_3-p_1q_3)+(q_6p_4-p_6q_4)\,,\qquad
M_{34}=(q_2p_3-p_2q_3)+(q_4p_5-p_4q_5)\,,
\end{align*}
are related to double rotations in $\mathbb R^6$, whereas functions
\begin{align*}
T_1=p_1^2 + p_2^2 + p_3^2\,,\qquad T_2=p_1^2 + p_4^2 + p_5^2\,,\\
T_3=p_2^2 + p_4^2 + p_6^2\,,\qquad T_4=p_3^2 + p_5^2 + p_6^2\,,
\end{align*}
are defined by triple translations. The direct calculations show that the Haantjes torsion of the corresponding Killing tensors is not zero.

From the sixteen integrals of motion $F_i, G_i$, $S_i$ and $W_i$ we have to choose six independent integrals, four of which are quadratic polynomials in momenta.

\section{Symmetric spaces of BD.I type}
\setcounter{equation}{0}
Symmetric space
\[\frac{SO(m+n)}{SO(m)\times SO(n)}\]
is only Hermitian when $m =2$ since in general $so(m) + so(n)$ has no centre. When $m = 2$ the $so (2)$ subalgebra is the centre and depending upon whether
$q$ is odd or even this symmetric space is associated with either $B_{(n+ 1)/2}$ or $D_{(n+ 2)/2}$ root systems.

The simplest nontrivial example is associated with $D_3$, and, similar to \cite{f83}, we present the Lax matrix for this system even though $D_3\cong A_3$.
\subsection{Euclidean space $\mathbb R^4$, case $m=n=2$}
We present $6\times 6$ Lax matrix (\ref{lax-gen}) using the same Cartan-Weil basis as in \cite{f83}
\[
L(\lambda)=\left(
\begin{matrix}
\widetilde{L}_{11}&\widetilde{L}_{12}\\
 \widetilde{L}_{21}&\widetilde{L}_{22}
\end{matrix}\right)\,.
\]
where
\begin{align*}
\widetilde{L}_{11}&=
\left(
 \begin{smallmatrix}
 -2 \lambda^2+2 q_1^2 + 2 q_2^2 + 2 q_3^2 + 2 q_4^2 + a_1 & p_1-2\mathrm i \lambda q_1 & p_2-2\mathrm i \lambda q_2 \\ \\
 p_1+2\mathrm i \lambda q_1 & -2 q_1^2 + 2 q_3^2 + a_2 & -2 q_1 q_2 + 2 q_3 q_4 \\ \\
 p_2+2\mathrm i \lambda q_2 & -2 q_1 q_2 + 2 q_3 q_4 & -2 q_2^2 + 2 q_4^2 + a_3 \\
 \end{smallmatrix}
 \right)
\\
\\
\widetilde{L}_{22}&=
\left(
 \begin{smallmatrix}
 \phantom{-}2 \lambda^2 - 2 q_1^2 - 2 q_2^2 - 2 q_3^2 - 2 q_4^2 + b_1 & -p_1-2\mathrm i \lambda q_1 & -p_2-2\mathrm i \lambda q_2 \\ \\
 -p_1+ 2\mathrm i \lambda q_1 & 2 q_1^2 - 2 q_3^2 + b_2 & 2 q_1 q_2 - 2 q_3 q_4 \\ \\
 -p_2+ 2\mathrm i \lambda q_2 & 2 q_1 q_2 - 2 q_3 q_4 & 2 q_2^2 - 2 q_4^2 + b_3 \\
 \end{smallmatrix}
 \right)
\end{align*}
and
\[\widetilde{L}_{12}=\left(
 \begin{smallmatrix}
 0 & p_3-2\mathrm i \lambda q_3 & p_4-2\mathrm i \lambda q_4 \\ \\
 -p_3+2\mathrm i \lambda q_3 & 0 & -2q_1q_4 + 2q_2q_3\\ \\
 -p_4+2\mathrm i \lambda q_4 & 2q_1q_4 - 2q_2q_3 & 0 \\
 \end{smallmatrix}
 \right)\,,\quad
\widetilde{L}_{21}=\left(
 \begin{smallmatrix}
 0 & -p_3-2\mathrm i \lambda q_3 & -p_4-2\mathrm i \lambda q_4 \\ \\
 p_3+2\mathrm i \lambda q_3 & 0 & 2q_1q_4 - 2q_2q_3\\ \\
 p_4+2\mathrm i \lambda q_4 & -2q_1q_4 + 2q_2q_3 & 0 \\
 \end{smallmatrix}
 \right) \,.
\]
Parameters satisfy the following relations
\[
a_2 = a_1 + b_1 - b_2\,,\qquad a_3 = a_1 + b_1 - b_3\,.
\]
In this case Hamiltonian $H$ (\ref{ham-gena}) has the form
\begin{align*}
H&=p_1^2 +p_2^2 +p_3^2+p_4^2+4(q_1^2+q_2^2+q_3^2+q_4^2)^2
 - 8(q_1q_3 + q_2q_4)^2\\\\
 &+ 2(b_2-b_1)q_1^2 +2 (b_3-b_1)q_2^2 + 2(a_1 - b_2)q_3^2 + 2(a_1-b_3)q_4^2
\end{align*}
Because $D_3\cong A_3$ this Hamiltonian coincides with (\ref{ham-13b}) up to rescaling and canonical transformation $q_i\to-q_i$ and $p_i\to-p_i$  of one of the coordinates and momenta.

The corresponding second-order Killing tensors are discussed in Section 2.

\subsection{Euclidean space $R^{2n-1}$, $m=2$.}
Let us consider representation of the Lie algebra $so(2n+1)$ by $(2n+1)\times (2n+1)$ matrices $X$ \cite{helg01}, which satisfy
\[X+SX^T S^{-1}=0\,,\qquad S=\sum_{k=1}^{2n+1} (-1)^{k+1}E_{k,2n+2-k}\,,\]
where $E_{ij}$ are matrices whose only non-zero entry is a unit in row $i$ and column $j$.

In this case, Cartan involution is related to the following element $\mathcal A= E_{1,1}-E_{2n+1,2n+1}$ of the Cartan subalgebra. In this representation Lax matrix (\ref{lax-gen}) has the following block structure
\[L(\lambda)=
\left(
  \begin{array}{ccc}
    2\lambda^2 & \vec{x}^T& 0 \\ \\
    \vec{y} &0& s\cdot\vec{x} \\ \\
    0 & \vec{y}^T\cdot s &- 2\lambda^2 \\
  \end{array}
\right)
+C+\Lambda\,,\]
where the central block of zeroes has dimensionality $(2n-1)\times (2n-1)$, the column vectors $x$ and $y$ have the following entries
\[ \vec{x}_i=p_i-2\mathrm i q_i\,,\quad \vec{y}_i=p_i+2\mathrm i q_i\,,\quad i=1,\ldots,2n-1\,,
\]
and $s$ is $(2n-1)\times (2n-1)$ matrix
\[
s=\sum_{k=1}^{2n-1} (-1)^k E_{k, 2n - k}\,.
\]
Matrix $\Lambda$ is a  numerical matrix which satisfies $\Lambda+S\Lambda^T S^{-1}=0$ and, following \cite{r86}, which determines a shift of the orbit.

\subsection{Euclidean space $\mathbb R^3$, case $n=3$}
For symmetric space $ \frac{SO(6)}{SO(2)\times SO(4)}$ we have the following $5\times 5$ Lax matrix
\[
L(\lambda)=\left(\begin{smallmatrix}
                 2\lambda^2 & p_1-2\mathrm i \lambda q_1 & p_2-2\mathrm i \lambda q_2 & p_3-2\mathrm i \lambda q_3 & 0 \\
                 p_1+2\mathrm i \lambda q_1 & 0 & 0 & 0 & -p_3+2\mathrm i \lambda q_3 \\
                 p_2+2\mathrm i \lambda q_2 & 0 & 0 & 0 & p_2-2\mathrm i \lambda q_2  \\
                 p_3+2\mathrm i \lambda q_3 & 0 & 0 & 0 & -p_1+2\mathrm i \lambda q_1 \\
                 0 & -p_3-2\mathrm i \lambda q_3 & p_2+2\mathrm i \lambda q_2 & -p_1-2\mathrm i \lambda q_1 & -2\lambda^2 \\
               \end{smallmatrix}\right)+2C+2\Lambda
\]
where
\[
C=\left(
    \begin{smallmatrix}
      -q_1^2 - q_2^2 - q_3^2 & 0 & 0 & 0 & 0 \\
      0 & q_1^2 - q_3^2 & (q_1 + q_3)q_2 & 0 & 0 \\
      0 & (q_1 + q_3)q_2 & 0 & (q_1 + q_3)q_2 & 0 \\
      0 & 0 & (q_1 + q_3)q_2 & -q_1^2 + q_3^2 & 0 \\
      0 & 0 & 0 & 0 & q_1^2 + q_2^2 + q_3^2 \\
    \end{smallmatrix}
  \right)\,,\quad
 \Lambda=\left(
           \begin{smallmatrix}
             a_1 & 0 & 0 & 0 & 0 \\
             0 & a_2 & a_3 & 0 & 0 \\
             0 & a_3 & 0 & a_3 & 0 \\
             0 & 0 & a_3 & -a_2 & 0 \\
             0 & 0 & 0 & 0 & -a_1 \\
           \end{smallmatrix}
         \right)\,.
\]
The Hamiltonian (\ref{ham-gen}) looks like
\begin{align*}
H&=\left.\frac14 \mbox{tr} L^2\right|_{\lambda=0}-2a_1^2 - 2a_2^2 - 4a_3^2=p_1^2 + p_2^2 + p_3^2 + 4(q_1^2 + q_2^2 + q_3^2)^2 - 2(2q_1q_3 - q_2^2)^2 \\ \\
&-4(a_1 - a_2)q_1^2 - 4(a_1 + a_2)q_3^2 - 4q_2\left(a_1q_2 - 2a_3(q_1 +q_3)\right)\,.
\end{align*}
The quadratic integral of motion
\[
F=(q_1p_2-p_1q_2+q_2p_3-q_3p_2)^2 -(p_1+p_3)(a_2(p_1 - p_3) + 2a_3p_2) +U
\]
where
\begin{align*}
U&=4(q_1 + q_3)\Bigl(a_2 (q_1 - q_3) + 2 a_3 q_2\Bigr) (a_1-q_1^2 - q_2^2 - q_3^2) - 4 (a_2^2 + a_3^2) (q_1^2 + q_3^2)\\ \\
& - 8 q_2 (q_1 - q_3) a_2a_3- 8 (q_1 q_3 + q_2^2) a_3^2\,,
\end{align*}
defines second-order Killing tensor with non-zero torsion.

The spectral curve of the Lax matrix is defined through the equation
\begin{align*}
&z^5 - 2(2\lambda^4 + 4a_1\lambda^2 + 2a_1^2 + 2a_2^2 + 4a_3^2 + H)z^3 \\
\\ &\qquad\qquad\qquad + \Bigl(16(a_2^2 + 2a_3^2)\lambda^4 + 8\bigl(F + 4a_1(a_2^2 + 2a_3^2)\bigr)\lambda^2 + G/2 - H^2\Bigr)z=0\,.
\end{align*}
The leading term of the polynomial of fourth order in momenta $G$ is defined by the curvatures tensor $\mathcal R$
\begin{align*}
G=&-\frac14\sum_{\alpha,\beta,\gamma,\delta} \mathcal R_{-\alpha,\beta,\gamma,-\delta}q^\alpha q^\beta q^\gamma q^\delta+\cdots\\ \\
&=4(p_1^2 + p_2^2 + p_3^2)^2 - 2(2p_1p_3 - p_2^2)^2+\cdots.
\end{align*}
When $a_i=0$ we have Hamiltonian (\ref{ham-13c}-\ref{c-pot}) up to canonical transformation
\[
H=\left.\frac14 \mbox{tr} L^2\right|_{\lambda=0}=p_1^2 + p_2^2 + p_3^2 + 4(q_1^2 + q_2^2 + q_3^2)^2 - 2(2q_1q_3 - q_2^2)^2\,,
\]
that follows from the equivalence of the symmetric spaces, see \cite{helg01}.
\subsection{Euclidean space $\mathbb R^5$, case $n=5$}
The $7\times 7$ Lax matrix reads as
\begin{align}
L(\lambda)&=\left(\begin{smallmatrix}
                 2\lambda^2 & p_1-2\mathrm i \lambda q_1 & p_2-2\mathrm i \lambda q_2 & p_3-2\mathrm i \lambda q_3 &  p_4-2\mathrm i \lambda q_4& \phantom{-}p_5-2\mathrm i \lambda q_5&0   \\
                 p_1+2\mathrm i \lambda q_1 & 0 & 0 & 0 &0 &0 &-p_5 +2\mathrm i \lambda q_5 \\
                 p_2+2\mathrm i \lambda q_2 & 0 & 0 & 0 &0 &0 &\phantom{-}p_4 -2\mathrm i \lambda q_4\\
                 p_3+2\mathrm i \lambda q_3 & 0 & 0 & 0 &0 &0 &-p_3+ 2\mathrm i \lambda q_3\\
                 p_4+2\mathrm i \lambda q_4 & 0 & 0 & 0 &0 &0 &\phantom{-}p_2-2\mathrm i \lambda q_2\\
                 p_5+2\mathrm i \lambda q_5 & 0 & 0 & 0 &0 &0 & -p_1+2\mathrm i \lambda q_1\\
                 0 & -p_5-2\mathrm i \lambda q_5&p_4+2\mathrm i \lambda q_4&-p_3-2\mathrm i \lambda q_3& p_2+2\mathrm i \lambda q_2 & -p_1-2\mathrm i \lambda q_1 & -2\lambda^2 \\
               \end{smallmatrix}\right)
\nn\\ \label{lax55}\\
&+2\left(
      \begin{smallmatrix}
        -\sum_{k=1}^5 q_k^2 & 0 & 0 & 0 & 0 & 0 & 0\\
        0 & q_1^2 - q_5^2 & q_1q_2 + q_4q_5 & q_3(q_1 - q_5) & q_1q_4 + q_2q_5 & 0 & 0 \\
        0 & q_1q_2 + q_4q_5  & q_2^2 - q_4^2 & q_3(q_2 + q_4) & 0 & q_1q_4 + q_2q_5  & 0 \\
        0 & q_3(q_1 - q_5) & q_3(q_2 + q_4) & 0 & q_3(q_2 + q_4) & -q_3(q_1 - q_5) & 0 \\
        0 & q_1q_4 + q_2q_5 & 0 & q_3(q_2 + q_4) & -q_2^2 +q_4^2 & q_1q_2 + q_4q_5 & 0 \\
        0 & 0 & q_1q_4 + q_2q_5 & -q_3(q_1 - q_5) &q_1q_2 + q_4q_5 &  -q_1^2 + q_5^2 & 0 \\
        0 & 0 & 0 & 0 & 0 & 0 & \sum_{k=1}^5 q_k^2 \\
      \end{smallmatrix}
    \right)\,.\nn
\end{align}
The corresponding Hamiltonian
\bq\label{ham-d5}
H=\left.\frac14 \mbox{tr} L^2\right|_{\lambda=0}=\sum_{k=1}^{5}p_k^2+4\left(\sum_{k=1}^{5}q_k^2\right)^2-2(2q_1q_5 - 2q_2q_4 + q_3^2)^2
\eq
commutes with the four linear integrals of motion
\begin{align*}
r_1&= (q_1p_2-p_1 q_2)+(q_4p_5-p_4q_5)\,,\qquad  r_2=(q_2p_3-p_2q_3) +(q_3p_4-p_3q_4)\,,\\ \\
r_3&= (q_1p_3-p_1q_3) + (q_5p_3-p_5q_3)\,,\qquad r_4 = (q_1p_4-p_1q_4)+(q_2p_5-p_2q_5)\,,
\end{align*}
so that
\begin{align*}
\{r_1, r_2\}&= - r_3\,,\quad \{r_1,r_3\}=r_2\,,\quad \{r_1,r_4\}=0\,,\\
\\
\{r_4, r_2\}&=  r_3\,,\quad \{r_4,r_3\}=-r_2\,,\quad \{r_2,r_3\}=r_4-r_1\,.
\end{align*}
The spectral curve of the Lax matrix
\[
\det(z\cdot I-L(\lambda))=z^7 -2(2\lambda^4 + H)z^5 + (8F_1\lambda^2 + G_1)z^3 - 4G_2z=0
\]
gives rise to four independent integrals of motion in involution $H$, $G_1$ and
\[F_1=(r_1^2 + r_2^2 + r_3^2 + r_4^2)\,,\qquad G_2=(r_1 + r_4)^2\Bigl((r_1 - r_4)^2 + 2(r_2^2 + r_3^2)\Bigr)\,.\]
Using Hamiltonian and fourth order polynomial $G_1$ we can get the integral of motion defined by a curvature tensor $\mathcal R$ (\ref{ham-gen}, \ref{ham-13b})
\begin{align*}
G_3=2G_1+2H^2&=-\frac14\sum_{\alpha,\beta,\gamma,\delta} \mathcal R_{-\alpha,\beta,\gamma,-\delta}p^\alpha p^\beta p^\gamma p^\delta+\cdots \\ \\
&=-4(p_1^2+p_2^2+p_3^3+p_4^2+p_5^2)^2 + 2(2p_1p_5 - 2p_2p_4 + p_3^2)^2+\cdots.
\end{align*}
which is independent on $H$ and $r_k$.

Because $\{r_k,G_1\}=0$ we have a completely integrable system with the five independent integrals of motion in involution, for instance
\[
r_1,\,r_4,\,r_2^2+r_3^2\,,H,\,G_1\,.
\]
Nevertheless, the Lax matrix (\ref{lax55}) generates only four of them, similar to the generalized Toda lattice \cite{deft86}.

By adding a constant matrix $\Lambda$ to $L(\lambda)$ (\ref{lax55}), where
\[
\Lambda=\left(
  \begin{array}{ccccccc}
    a_1 & 0 & 0 & 0 & 0 & 0 & 0 \\
    0 & a_2 & 0 & 0 & 0 & 0 & 0 \\
    0 & 0 & a_3 & a_4 & 0 & 0 & 0 \\
    0 & 0 & a_4 & 0 & a_4 & 0 & 0 \\
    0 & 0 & 0 & a_4 & -a_3 & 0 & 0 \\
    0 & 0 & 0 & 0 & 0 & -a_2 & 0 \\
    0 & 0 & 0 & 0 & 0 & 0 & -a_1 \\
  \end{array}
\right)\,,
\]
we have
\begin{align*}
H=&\left.\frac14 \mbox{tr} L^2\right|_{\lambda=0}-2a_1^2 - 2a_2^2 - 2a_3^2 - 4a_4^2
=\sum_{k=1}^{5}p_k^2+4\left(\sum_{k=1}^{5}q_k^2\right)^2-2(2q_1q_5 - 2q_2q_4 + q_3^2)^2\\ \\
+&(a_1 - a_2)q_1^2 + (a_1 - a_3)q_2^2 +q_3(a_1q_3 - 2a_4q_2 - 2a_4q_4)+ (a_1+ a_3)q_4^2 + (a_2 + a_1)q_5^2\,.
\end{align*}
In this case equation for the spectral curve
\begin{align*}
z^7 - 4(\lambda^4 - 2\lambda^2a_1 +a_1^2+a_2^2+a_3^2 + 2a_4^2 +H/2)z^5 +& \Bigl(16(a_2^2 + a_3^2 + 2a_4^2)\lambda^4  + F_1\lambda^2 + G_1\Bigr)z^3\\
\\
 -&\Bigl(64a_2^2(a_3^2 + 2a_4^2)\lambda^4 + F_2\lambda^2 + G_2\Bigr)z=0
\end{align*}
contains a sufficient number of integrals of motion for integrability by the Liouville theorem. There are three polynomials $H,F_1,F_2$ of the second order in momenta and two polynomials $G_1$ and $G_2$ of the fourth order in momenta.

\section{Reductive Homogeneous Spaces}
According to \cite{f83} in previous sections, we consider symmetric spaces which are reductive homogeneous spaces on which the canonical
connections have zero torsion. In this section, we consider one example associated with reductive homogeneous spaces which have non-zero torsion.

Following \cite{f83} let us consider symmetric space
\[
\frac{SU(3)}{S(U(1) \times U(1) \times U(1))}
\]
and $3\times 3$ Lax matrix
\[L=\left(
 \begin{array}{ccc}
 a_1\lambda^2 + 2d(q_1^2 + q_2^2) + w_1 & b\lambda q_1 +p_1+ 2d q_2q_3&b\lambda q_2 + p_2  \\ \\
 -b\lambda q_1 +p_1+ 2dq_2q_3  & a_2\lambda^2 -2d(q_1^2-q_3^2) + w_2 & b\lambda q_3+p_3 - 2dq_1q_2 \\ \\
-b\lambda q_2 + p_2  & -b\lambda q_3+p_3 - 2dq_1q_2 & a_3\lambda^2 - 2d(q_2^2 - q_3^2) + w_3 
 \end{array}
 \right)\,.
\]
where  we have to fix eigenvalues of $\mathcal A$  in a special way 
\[
a_2=a_3=-\frac{a_1}{2}\,,
\]
put $b = \sqrt{3 a_1 d\,}$ and $q_3=0$, i.e. make $q_3$ cyclic coordinate \cite{f83}. 

The corresponding Hamiltonian 
\begin{align*}
H=&\left.\frac14 \mbox{tr} L^2\right|_{\lambda=0}-\frac{1}{4}\sum \omega_i^2-\omega_2^2=\frac{p_1^2+p_2^2+p_3^2}{2}+2d^2(q_1^2 + q_2^2)^2
\\
 -& 2dq_1q_2p_3 + d(w_1 - w_2)q_1^2 + d(w_1 - w_3)q_2^2\,,
\end{align*}
commutes with the linear integral of motion $p_3$ and with the following cubic integral of motion 
\begin{align*}
K_3&=2p_1p_2p_3-w_1p_3^2 - w_2p_2^2 - w_3p_1^2-4(q_1^2 + q_2^2)(w_2q_2^2 + w_3q_1^2 - 2p_3q_1q_2)d^2\\
\\
&+2d\bigl((q_1p_2 -  q_2p_1)^2-(q_1^2 + q_2^2)p_3^2\bigr)
-2d\bigl(w_3(w_1 - w_2)q_1^2 + w_2(w_1 - w_3)q_2^2\bigr)
+4dw_1p_3q_1q_2\,,
\end{align*}
so that
\[
\left.\frac13 \mbox{tr} L^3\right|_{\lambda=0}=K_3+2(w_1 + w_2 + w_3)H+const\,.
\]
If  we put $p_3=const$ we obtain two quadratic integrals of motion. 
Other Lax matrices associated with the three-wave interaction system are discussed in \cite{kost08}.

\section{Conclusion}
We present examples of the Killing tensors of valency two generating quadratic integrals of motion for the integrable systems having additional integrals of motion which are polynomials of higher order in momenta.

All these Killing tensors are related to the special combinations of rotations and translations. It will be interesting to find criteria which allow us to extract these special Killing tensors from the generic solution of the Killing equation on Euclidean, Riemannian and pseudo-Riemannian spaces of constant curvature.

\vskip0.2truecm
The work was supported by the Russian Science Foundation (project 21-11-00141).

\vskip0.2truecm

The second author (AVT) gratefully acknowledges the kind hospitality provided by
Yanqi Lake Beijing Institute of Mathematical Sciences and Applications during his stay in Fall 2022 when work on this text was finished.

\end{document}